\documentclass[12pt]{article}
\usepackage{amsmath}
\usepackage{amssymb}
\usepackage{mathtools}
\usepackage{siunitx}
\usepackage{graphicx,psfrag,epsf}
\usepackage{url}
\usepackage[dvipsnames]{xcolor}
\usepackage{enumerate}
\usepackage[shortlabels]{enumitem}
\usepackage{cancel}
\usepackage{soul}
\usepackage{dsfont}
\usepackage{multirow}
\RequirePackage[authoryear]{natbib}

\RequirePackage{amsthm,amsmath,amsfonts,amssymb}
\newtheorem{theorem}{Theorem}
\newtheorem{proposition}{Proposition}
\RequirePackage[colorlinks,citecolor=Greatcolor,urlcolor=blue]{hyperref}
\RequirePackage{graphicx}
\usepackage{caption}
\usepackage{subcaption}
\usepackage{float}
\usepackage{url}
\usepackage{bbm}
\usepackage{booktabs,caption,subcaption}
\definecolor{MSBlue}{rgb}{.204,.353,.541}
\definecolor{MSLightBlue}{rgb}{.31,.506,.741}
\definecolor{WeiBlue}{rgb}{ .184,  .459,  .71}
\definecolor{tabcolor}{rgb}{ .357,  .608,  .835}
\definecolor{codegreen}{rgb}{0,0.6,0}
\definecolor{codegray}{rgb}{0.5,0.5,0.5}
\definecolor{codepurple}{rgb}{0.58,0,0.82}
\definecolor{backcolour}{rgb}{0.95,0.95,0.92}
\definecolor{codecomment}{RGB}{77,137,109}
\definecolor{codestring}{RGB}{255,179,93}
\definecolor{codekey}{RGB}{6,63,255}
\definecolor{Greatcolor}{RGB}{228, 114,114}

\usepackage[margin=1in]{geometry}

\makeatletter
\newcommand*{\rom}[1]{\expandafter\@slowromancap\romannumeral #1@}
\makeatother

\DeclareMathOperator{\Cov}{Cov}
\DeclareMathOperator{\Corr}{Corr}

\newcommand{\E}{\mathbb{E}}

\DeclareMathOperator{\Var}{Var}

\newcommand{\bZ}{\boldsymbol{Z}}

\newcommand{\bbeta}{\boldsymbol{\beta}}

\newcommand{\bmu}{\boldsymbol{\mu}}
\newcommand{\bOmega}{\boldsymbol{\Omega}}

\newcommand{\bSigma}{\boldsymbol{\Sigma}}

\newcommand{\btheta}{\boldsymbol{\theta}}

\usepackage{authblk}
\usepackage{setspace}
\usepackage{xr-hyper}

\date{}
\title{Estimating Covariate Effects on Functional Connectivity using Voxel-Level fMRI Data}

\author[1]{Wei Zhao\thanks{\texttt{wzhao24@ncsu.edu}}}
\author[1]{Brian J. Reich\thanks{\texttt{brian\_reich@ncsu.edu}}}
\author[1,2]{Emily C. Hector\thanks{Corresponding author: \texttt{ehector@umich.edu}}}

\affil[1]{Department of Statistics, North Carolina State University, Raleigh, NC 27695, U.S.A.}
\affil[2]{Department of Biostatistics, University of Michigan, Ann Arbor, MI 48109, U.S.A.}

\begin{document}

\def\spacingset#1{\renewcommand{\baselinestretch}%
{#1}\small\normalsize} \spacingset{1}

\label{firstpage}

\maketitle
%  put the summary for your paper here

\begin{abstract}
Functional connectivity (FC) analysis of resting-state fMRI data provides a framework for characterizing brain networks and their association with participant-level covariates. Due to the high dimensionality of neuroimaging data, standard approaches often average signals within regions of interest (ROIs), which ignores the underlying spatiotemporal dependence among voxels and can lead to biased or inefficient inference. We propose to use a summary statistic---the empirical voxel-wise correlations between ROIs---and, crucially, model the complex covariance structure among these correlations through a new positive definite covariance function. Building on this foundation, we develop a computationally efficient two-step estimation procedure that enables statistical inference on covariate effects on region-level connectivity. Simulation studies show calibrated uncertainty quantification, and substantial gains in validity of the statistical inference over the standard averaging method. With data from the Autism Brain Imaging Data Exchange, we show that autism spectrum disorder is associated with altered FC between attention-related ROIs after adjusting for age and gender. The proposed framework offers an interpretable and statistically rigorous approach to estimation of covariate effects on FC suitable for large-scale neuroimaging studies.
\end{abstract}

%  Please place your key words in alphabetical order, separated
%  by semicolons, with the first letter of the first word capitalized,
%  and a period at the end of the list.
%

\noindent%
{\it Keywords:} Autism spectrum disorder; Covariance modeling; Isserlis's theorem; Resting-state fMRI; Spatiotemporal dependence.

%  As usual, the \maketitle command creates the title and author/affiliations
%  display 
\maketitle
\spacingset{1.45}
\section{Introduction}
\label{sec:intro}

Understanding the functional organization of the human brain through analysis of the dependence between activated brain regions of interest (ROIs), termed functional connectivity (FC), has become a cornerstone of modern neuroimaging research \citep{aertsen1989dynamics, friston1993functional, biswal1995functional, friston2011functional} and the development of individualized connectome-based biomarkers \citep{finn2015functional}. FC analysis has led to a range of scientific discoveries, including the identification of the default mode network \citep{raichle2001default}, and insights into altered brain connectivity in neuro developmental and neuropsychiatric disorders such as autism spectrum disorder (ASD) and schizophrenia \citep{kennedy2006failing, ramsay2017increases}. FC is typically measured using the sample correlations between resting-state functional Magnetic Resonance Imaging (rs-fMRI) time series, which measure the blood oxygenation in the brain and therefore serve as a proxy for brain activation, in anatomically distinct brain ROIs \citep{biswal1997simultaneous, lowe2000correlations, cordes2000mapping, greicius2003functional, van2010exploring}. 
Altered FC has been implicated in  neuropsychiatric and neurodevelopmental disorders such as ASD. Since the first publications using rs-fMRI in ASD research \citep{cherkassky2006functional, kennedy2006failing, kennedy2008intrinsic, hull2017resting}, however, findings have been inconsistent. Some studies report under-connectivity (also termed hypoconnectivity, decreased connectivity, or weaker connectivity) between frontal and posterior regions \citep{just2004cortical, just2012autism} or within the default mode network \citep{greicius2003functional, assaf2010abnormal}, while others report over-connectivity (also termed hyperconnectivity, increased connectivity, or stronger connectivity) \citep{supekar2013brain, cerliani2015increased}, and some show mixed patterns \citep{benkarim2021connectivity, hull2017resting, picci2016theoretical}. These inconsistencies likely stem from a combination of biological heterogeneity in ASD and methodological variations across studies. Participant-level factors such as age, gender, IQ, and race are often unevenly controlled, introducing potential confounding. Analytical differences, including the use of ROI-based methods or whole-brain approaches, inconsistent multiple comparison correction, and variability in preprocessing pipelines further contribute to divergent findings \citep{halliday2024heterogeneity}. This variability underscores the need for statistical methods that can flexibly model participant-level variability and quantify uncertainty in estimated connectivity while controlling Type-I error rates.

A variety of statistical methods have been proposed to model FC from rs-fMRI data \citep{hull2017resting, van2010exploring}. One of the most widely used model-based approaches for exploring FC is the ROI seed-based method. This technique defines a seed region based on prior literature and computes the correlation between its average time series and those of other voxels or ROIs, resulting in a FC map measuring the FC between the seed and the rest of the brain \citep{cerliani2015increased, di2011aberrant, weng2010alterations}. While straightforward to interpret, seed-based methods are limited to connectivity patterns involving the selected seed. To enable whole-brain analyses, mass univariate models fit linear models independently to each Fisher-transformed sample correlation of ROI-averaged rs-fMRI time series, allowing for covariate adjustment but ignoring dependence between connections. The multiple tests of significance inflate the Type-I error rate, requiring multiple comparison corrections that are overly conservative in the presence of spatiotemporal dependence \citep{benjamini1995controlling, chumbley2009false}. Multivariate distance matrix regression (MDMR) \citep{shehzad2014multivariate} provides a nonparametric alternative to these two approaches by testing associations between whole-brain connectivity patterns and covariates using distance-based statistics and permutation tests. While MDMR enables voxel- or ROI-wise testing without strong distributional assumptions, it does not explicitly account for the correlation between connections (i.e. sample correlations) and typically relies on a multiple comparison correction due to the mass univariate testing, which can be overly conservative and lead to reduced power \citep{ponsoda2017structural}. 

More recent model-based methods address these limitations by jointly estimating FC between ROIs. \cite{kim2023bayesian} propose a Bayesian clustering framework that partitions high-dimensional covariance matrices into block-diagonal structures, with each block corresponding to a group of correlated ROIs. This approach flexibly captures ROI-level dependence patterns but is not designed to adjust for covariates. \cite{kim2025new} introduce a block covariance regression model that estimates participant-specific voxel-level connectivity while adjusting for covariates in a Bayesian framework. Their method leverages the block structure induced by defined ROIs for model parsimony and uses a modified Cholesky decomposition for scalable inference, but the covariate model can be difficult to interpret. \cite{castruccio2018scalable} develop a multi-resolution spatiotemporal model that accounts for spatial anisotropy within ROIs using nonstationary Gaussian processes, models temporal dependence via voxel-specific vector auto-regressive processes, and estimates connectivity between ROIs via a graphical Least Absolute Shrinkage and Selection Operator. However, this approach does not incorporate subject-level covariates. There remains a need for methods that directly model covariate-dependent FC between ROIs using the rich voxel-level rs-fMRI time series, while simultaneously accounting for spatiotemporal covariance among voxel-level time series in a computationally scalable framework. This need is particularly motivated by our interest in understanding how ASD affects FC between brain ROIs.

We propose a computationally efficient framework for ROI-level FC regression on participant covariates while accounting for voxel-level spatiotemporal dependence. Unlike traditional methods that either ignore voxel-level structure or assume independence between correlations, we directly model the complex covariance structure between the empirical voxel-wise correlations between ROIs, which arises from the spatiotemporal dependence in voxel-level rs-fMRI time series. We derive the asymptotic joint distribution of the between-ROI empirical correlations under a hierarchical spatiotemporal model that adjusts for covariate effects, and prove that its covariance is positive definite. This contribution addresses a key barrier in FC modeling: capturing the complex covariance among correlation estimates themselves, which is critical for accurate and efficient uncertainty quantification and hypothesis testing that bypasses conservative multiple comparison adjustments. 

On this foundation, we develop a two-step estimation procedure that separates voxel-level covariance estimation from participant-level FC regression. This modular design dramatically reduces computational cost relative to full likelihood methods, while enabling asymptotically valid inference on the association between covariates and ROI-level connectivity. To our knowledge, our framework is the first to (i) establish the asymptotic distribution of empirical voxel-wise correlations, (ii) derive their complex dependence structure analytically, and (iii) incorporate this into a computationally scalable inference framework for participant-level analysis. 

The remainder of this paper is organized as follows. Section \ref{sec:method} introduces the proposed statistical model and two-step estimation framework. Section \ref{sec:simulation} presents simulation studies and Section \ref{sec:application} investigates the FC of ASD with data from the Autism Brain Imaging Data Exchange (ABIDE). Section \ref{sec:discussion} concludes with a discussion of future directions. All proofs are deferred to the supplement.

\section{Method}\label{sec:method}

We describe the proposed statistical framework enabling ROI-level FC regression while accounting for voxel-level spatiotemporal dependence. The method builds on a summary statistic---the empirical correlation between voxels in different ROIs---and explicitly models the complex covariance structure among these correlations induced by spatial and temporal dependencies in the rs-fMRI time series. Leveraging this structure, we derive the asymptotic distribution of the empirical correlation vector and develop a computationally efficient two-step estimation procedure for covariate effects. This framework facilitates scalable inference on ROI-level connectivity and its association with participant-level covariates.

\subsection{Voxel-Level Spatiotemporal Model}
\label{sec:model}
Let $i = 1,\dots, N$ index independent participants, $\text{ROI}_k, k \in \{1,2\}$ denote two anatomically distinct brain ROIs, each comprised of $n_k$ voxels. For simplicity of exposition, we consider two ROIs but our method can be generalized to more than two ROIs. Let $\mathbf{X}_i \in \mathbb{R}^p$ denote the covariate vector for participant $i$, consisting for example of an intercept, autism status, age and gender. Denote by $Z_{ikt}(\mathbf{s})$ the observed rs-fMRI signal at voxel location $\mathbf{s} \in \text{ROI}_k$, a 3D voxel spatial coordinate in region $k$, for participant $i$ at time $t \in \{ 1,\dots, T_i\}$. As is standard in the rs-fMRI literature \citep{power2014methods, carre2020standardization, wahid2021intensity}, we assume that, for each participant and ROI, the observed signals $Z_{ikt}(\mathbf{s})$ are centered and scaled to have zero mean and unit variance across all voxels and time points. We further assume that the rs-fMRI signal at each voxel can be decomposed as
\begin{equation}\label{eq:model}
  Z_{ikt}(\mathbf{s}) = a_{ikt} + u_{ikt}(\mathbf{s}) + e_{ikt}(\mathbf{s}),  
\end{equation}
where $a_{ikt}$ is the ROI-level signal shared across voxels within region $\text{ROI}_k$, $u_{ikt}(\mathbf{s})$ is the voxel-level spatiotemporal signal, and $e_{ikt}(\mathbf{s})$ is measurement error. We assume that $a_{ikt}, u_{ikt}$ and $e_{ikt}$ are mutually independent with distributions described next.

We assume $(a_{i1t},a_{i2t})$ are independent across $i$ and $t$ such that
$$
(a_{i1t},a_{i2t})^\top \sim \mathcal{N}\Biggl\{\mathbf{0}, \Biggl( \begin{array}{cc}
    \lambda_1^2 & \lambda_1 \lambda_2 \rho_i\\
    \lambda_1 \lambda_2 \rho_i &  \lambda_2^2
\end{array} \Biggr) \Biggr\},
$$
where $\lambda_k^2 >0$ denotes the variance of the ROI-level signal $a_{ikt}$ in $\text{ROI}_k$, and $\rho_i \in (-1,1)$ represents the participant-specific across-region connectivity. The connectivity $\rho_i$ depends on participant-level covariates through
$$\mbox{logit}\{ (\rho_i+1)/2 \} = \mathbf{X}_i^\top\boldsymbol{\beta},$$
where $\mbox{logit}(x)=\log\{x/(1-x)\}$ and $\boldsymbol{\beta} \in \mathbb{R}^p$ is the participant-level covariate effect on $\rho_i$. The vector $\boldsymbol{\beta}$ serves as the primary parameter of interest. 

We assume that $u_{ikt}(\mathbf{s})$ is independent across $i$ and $k$, and follows a zero-mean Gaussian process with a separable spatiotemporal covariance structure
$$
\Cov\{u_{ikt}(\mathbf{s}),u_{ikt'}(\mathbf{s}')\} = \sigma_k^2r_{s_k}(\mathbf{s},\mathbf{s}')r_\tau(t,t'),
$$
where $\sigma_k^2 > 0$ denotes the marginal variance of $u_{ikt}(\mathbf{s})$,  $r_{s_k}(\mathbf{s},\mathbf{s}') = \Corr\{u_{ikt}(\mathbf{s}), u_{ikt}(\mathbf{s}')\}$ denotes the spatial correlation between two locations $\mathbf{s}$ and $\mathbf{s}'$ in $\text{ROI}_k$, and $r_\tau(t,t') = \Corr\{u_{ikt}(\mathbf{s}),u_{ikt'}(\mathbf{s})\}$ denotes the temporal correlation. While other choices are possible, we adopt the exponential spatial kernel $r_{s_k}(\mathbf{s},\mathbf{s}') = \exp(-\|\mathbf{s} - \mathbf{s}'\|/\psi_k)$ with spatial range parameter $\psi_k > 0$, and an AR(1) temporal structure $r_\tau(t,t') = \phi^{|t - t'|}$ with $\phi \in (0,1]$ governing the temporal correlation. While we focus on $\phi > 0$ due to the nature of the rs-fMRI time series, the model can be easily extended to allow negative values of $\phi$.

We assume that the measurement error $e_{ikt}(\mathbf{s}) \sim \mathcal{N}(0,\tau_k^2)$ independently and identically across $k$, $\mathbf{s}$, $t$ and $i$. 
Due to the centering and scaling of the observed signals, the variance satisfies the constraint $\lambda_k^2 + \sigma_k^2 + \tau_k^2 = 1$ with $\lambda_k^2, \sigma_k^2, \tau_k^2 > 0$, $k = 1,2$. 

To express the model more compactly, we define $\mathbf{Z}_{ik} = \{Z_{ikt}(\mathbf{s})\}_{\mathbf{s}\in \text{ROI}_k,t = 1,\dots,T_i} \in \mathbb{R}^{n_kT_i}$ as the vector of observed voxel-level signals for participant $i$ in region $k$. The vectorized model is given by
$
\mathbf{Z}_{ik} = \mathbf{a}_{ik}\otimes \mathbf{1}_{n_k} + \mathbf{u}_{ik} + \mathbf{e}_{ik},
$
where $\mathbf{a}_{ik} = (a_{ik1}, \dots, a_{ikT_i})^\top \in \mathbb{R}^{T_i}$, $\mathbf{1}_{n_k} \in \mathbb{R}^{n_k}$ is a vector of ones and $\otimes$ denotes the Kronecker product. The terms $\mathbf{u}_{ik} = \{u_{ikt}(\mathbf{s})\}_{\mathbf{s}\in \text{ROI}_k,t = 1,\dots,T_i}$ and $\mathbf{e}_{ik} = \{e_{ikt}(\mathbf{s})\}_{\mathbf{s}\in \text{ROI}_k,t = 1,\dots,T_i} \in \mathbb{R}^{n_kT_i}$ represent the stacked voxel-level spatiotemporal process and measurement error for participant $i$ in region $k$. Under this construction, for each $k \in \{1,2\}$ the vector $\mathbf{Z}_{ik}$ follows a multivariate normal distribution $\mathbf{Z}_{ik} \sim \mathcal{N}(\mathbf{0},\bSigma(\btheta_k))$
where $\btheta_k = (\lambda_k^2, \sigma_k^2, \psi_k, \phi, \tau_k^2)^\top$ and
\begin{equation}
    \bSigma(\btheta_k) = \lambda_k^2 (\mathbf{I}_T \otimes \mathbf{J}_{n_k}) + \sigma_k^2 (\mathbf{R}_t \otimes \mathbf{R}_{s_k}) + \tau_k^2 \mathbf{I}_{n_kT},
    \label{eq:covz}
\end{equation}
where $\mathbf{J}_{n_k} \in \mathbb{R}^{n_k\times n_k}$ is a matrix of ones, and $\mathbf{I}_{T_i} \in \mathbb{R}^{T_i\times T_i}$ and $\mathbf{I}_{n_kT_i} \in \mathbb{R}^{(n_kT_i) \times (n_kT_i)}$ are identity matrices. The temporal correlation matrix $\mathbf{R}_t \in \mathbb{R}^{T_i\times T_i}$ captures temporal dependence across time points, with entries $[\mathbf{R}_t]_{ij} = r_\tau(i,j)$. The spatial correlation matrix $\mathbf{R}_{s_k} \in \mathbb{R}^{n_k \times n_k}$ encodes spatial correlations among voxel locations in region $k$, with entries $[\mathbf{R}_{s_k}]_{ij} = r_{s_k}(\mathbf{s}_i,\mathbf{s}_j)$. 

The proposed model is highly flexible as it allows for heterogeneity through the variance parameters $(\lambda_{1}^2, \lambda_{2}^2, \sigma_{1}^2, \sigma_{2}^2, \tau_{1}^2, \tau_{2}^2)$, spatial and temporal correlation structures ($\mathbf{R}_{s_1}, \mathbf{R}_{s_2}$ and $\mathbf{R}_{t}$), as well as participant-specific between ROI-correlation through $\rho_i$. This flexibility enables modeling of both participant-specific variability and heterogeneity in spatial and temporal dependence structures of rs-fMRI signals. %For computational tractability and a streamlined interpretation, however, we will often assume a homogeneous setting where spatiotemporal structures are shared across participants. 

\subsection{Proposed Summary Statistic: Voxel-wise Temporal Correlation Across ROIs}
Jointly modeling the voxel-level time series across two (or more) ROIs using a full likelihood approach is computationally infeasible in typical fMRI settings. For example, modeling two ROIs with $n_k=100$ voxels each and $T_i=200$ time points yields a spatiotemporal vector of length $(n_1+n_2)T_i=40{,}000$, leading to a $40{,}000 \times 40{,}000$ covariance matrix per participant. Extending this to multiple ROIs or whole-brain analysis, where the number of voxels often exceeds several thousands, leads to FC matrices with dimensions in the billions. Evaluating and inverting such high-dimensional structures for every participant is prohibitively expensive. 

To overcome this bottleneck, we introduce a low-dimensional summary: the empirical correlation between voxels in different ROIs across time. This reduces the dimensionality from $\mathcal{O}(n_1T_i + n_2T_i)$ to $\mathcal{O}(n_1n_2)$, which drastically accelerates computation while retaining voxel-level information about across-region connectivity. Specifically, for each participant, we define the empirical correlation vector as the collection of sample correlations of time series across all voxel pairs between two ROIs. For each voxel pair $(\mathbf{s}_1,\mathbf{s}_2)$ where $\mathbf{s}_1\in \text{ROI}_1$ and $\mathbf{s}_2\in \text{ROI}_2$, we compute
\begin{equation}\label{eq:voxelconnectivity}
  Y_i(\mathbf{s}_1,\mathbf{s}_2) = \frac{1}{T_i}\sum_{t = 1}^{T_i} Z_{i1t}(\mathbf{s}_1)Z_{i2t}(\mathbf{s}_2),  
\end{equation}
and stack these values into a vector $\mathbf{Y}_i=\{ Y_i(\mathbf{s}_1, \mathbf{s}_2) \}_{\mathbf{s_1} \in \text{ROI}_1, \mathbf{s}_2 \in \text{ROI}_2} \in \mathbb{R}^{n_1n_2}$.
% $$
% \mathbf{Y}_i = \left\{\frac{1}{T}\sum_{t = 1}^T Z_{i1t}(s_1)Z_{i2t}(s_2)\right\}_{s_1\in \text{ROI}_1,s_2\in \text{ROI}_2}.
% $$
Thus, $\mathbf{Y}_i$ provides a condensed summary of voxel-level connectivity between ROIs. We next derive the finite sample mean and variance and the asymptotic distribution of $\mathbf{Y}_i$ under the hierarchical model proposed in Section \ref{sec:model}. 
\begin{theorem}\label{thm:empirical_corr}
    Assume $Z_{ikt}(\mathbf{s})$ admits the model in \eqref{eq:model}. Let $\btheta = (\lambda_1^2, \lambda_2^2, \sigma_1^2, \sigma_2^2, \tau_1^2, \tau_2^2, \psi_1, \psi_2, \phi)^\top$ denote the shared spatiotemporal covariance parameters. Then the mean and the covariance of $\mathbf{Y}_i$ are given by $ \E(\mathbf{Y}_i)=\bmu_{\mathbf{Y}_i} =\rho_i\lambda_1\lambda_2\mathbf{1}_{n_1n_2}$, $\Var(\mathbf{Y}_i)=\bSigma_{\mathbf{Y}_i} = \{\rho_i^2\lambda_1^2\lambda_2^2\mathbf{J}_{n_1n_2} + \bOmega_i(\btheta)\}/T_i$, with $\bOmega_i(\btheta)$, which captures spatiotemporal variability, defined explicitly in equation \eqref{eq:OmegaFull} below, and its derivation is provided in Appendix \ref{appendix:proof_thm1}. Further, $\mathbf{Y}_i$ is asymptotically normal: \begin{equation}\label{eq:Yi_distribution}
\mathbf{Y}_i \stackrel{d}{\to} \mathcal{N} (\bmu_{\mathbf{Y}_i}, \bSigma_{\mathbf{Y}_i} ) \quad \text{as } T_i \to \infty.
\end{equation}
\end{theorem}
\noindent
The covariance of $\mathbf{Y}_i$, $\bSigma_{\mathbf{Y}_i}$, is derived using Isserlis's theorem \citep{isserlis1918formula}, which gives the form of fourth-order moments of Gaussian variables. 
The mean of $\mathbf{Y}_i$, $\bmu_{\mathbf{Y}_i}$, arises from the contribution of the ROI-level signals and is solely determined by the participant-specific connectivity parameter $\rho_i$, scaled by $\lambda_1\lambda_2$, and is constant across all voxel pairs. This implies that, marginally, voxel-level correlations between ROIs are expected to be homogeneous across voxel locations and fully characterized by the participant's connectivity strength. 

The covariance of $\mathbf{Y}_i$ reflects variability induced by both signal and noise components and is scaled by $1/T_i$ due to temporal averaging across $T_i$ time points in \eqref{eq:voxelconnectivity}. Specifically, the covariance consists of two components, each reflecting a different source of variation in the cross-ROI sample correlations. The first term, $\rho_i^2\lambda_1^2\lambda_2^2\mathbf{J}_{n_1n_2}$, corresponds to the contribution of the shared ROI-level signal. Because this signal is common across all voxels within a region, it induces a rank-one component in the covariance structure. The second term, $\bOmega_i(\btheta)$, accounts for the remaining variability due to voxel-specific spatiotemporal dependence and measurement error. The full expression for $\bOmega_i(\btheta)$ is
\begin{equation}
\begin{aligned}\label{eq:OmegaFull}
\bOmega_i(\btheta) = &\; 
\lambda_1^2 \lambda_2^2 \mathbf{J}_{n_1 n_2} 
+ \lambda_1^2 \sigma_2^2 (\mathbf{J}_{n_1} \otimes \mathbf{R}_{s_2})
+ \lambda_1^2 \tau_2^2 (\mathbf{J}_{n_1} \otimes \mathbf{I}_{n_2}) \\
&+ \lambda_2^2 \sigma_1^2 (\mathbf{R}_{s_1} \otimes \mathbf{J}_{n_2})
+ \sigma_1^2 \sigma_2^2 m(\phi,T_i) (\mathbf{R}_{s_1} \otimes \mathbf{R}_{s_2})
+ \sigma_1^2 \tau_2^2 (\mathbf{R}_{s_1} \otimes \mathbf{I}_{n_2}) \\
&+ \lambda_2^2 \tau_1^2 (\mathbf{I}_{n_1} \otimes \mathbf{J}_{n_2})
+ \sigma_2^2 \tau_1^2 (\mathbf{I}_{n_1} \otimes \mathbf{R}_{s_2})
+ \tau_1^2 \tau_2^2 (\mathbf{I}_{n_1} \otimes \mathbf{I}_{n_2}),
\end{aligned}
\end{equation}
where the function $m(\phi,T_i)$ accounts for the temporal autocorrelation and takes value $T_i$ when $\phi=1$ and value 
$$
%m(\phi,T_i)=
%\begin{cases}
%\displaystyle 
%T_i, 
%& \phi = 1,\\[10pt]
%\displaystyle 
1+\frac{2\phi^{2}\{1-\phi^{2(T_i-1)}\}}{1-\phi^{2}}
   -\frac{2\phi^{2}\{1-T_i \phi^{2(T_i-1)} + (T_i-1)\phi^{2T_i}\}}
          {T_i(1-\phi^{2})^{2}}
\;\xrightarrow[T_i\to\infty]{}\;
\dfrac{1+\phi^{2}}{1-\phi^{2}},
%& \phi \in (0,1],
%\end{cases}
$$
when $\phi \in (0,1)$.
A detailed derivation of the temporal scaling function $m(\phi,T_i)$ is provided in Appendix \ref{appendix:proof_thm1}. As $T_i \to \infty$ with $|\phi| < 1$, the covariance $\bSigma_{\mathbf{Y}_i}$ approaches zero, since the sample correlation $\mathbf{Y}_i$ defined in~\eqref{eq:voxelconnectivity} converges in probability to its mean $\bmu_{\mathbf{Y}_i}$, and the variability across voxel pairs vanishes. In the special case where $\phi = 1$, the temporal autocorrelation implies perfect dependence across all time points, reducing the temporal dimension to one effective sample. In this case, the covariance $\bSigma_{\mathbf{Y}_i}$ reflects purely spatial dependence. In practice, when $T_i$ is large but finite, all variance components remain non-negligible. If spatial dependence among voxels is ignored, the variability in the empirical correlation $\mathbf{Y}_i$ is underestimated, leading to inflated test statistics and Type I error rates. Therefore, for valid inference on ROI-level connectivity at finite $T_i$, especially when estimating covariate effects, it is essential to account for the dependence between sample correlations explicitly to ensure accurate uncertainty quantification and control false discoveries.

To guarantee the well-posedness of the model and the applicability of Gaussian approximation, the following proposition establishes that the covariance matrix $\bSigma_{\mathbf{Y}_i}$ is positive definite. 
\begin{proposition}\label{prop:pd}
Assume that the spatial covariance matrices $\mathbf{R}_{s_1}$ and $\mathbf{R}_{s_2}$ are positive definite, the marginal spatiotemporal variances $\sigma_1^2, \sigma_2^2$ are strictly positive, and the temporal autocorrelation parameter satisfies $\phi \in (0,1]$. Then the covariance matrix $\bSigma_{\mathbf{Y}_i}$ is positive definite.
\end{proposition}

\subsection{Two-Step Estimation Framework}\label{sec:step1}
We next outline a two-step estimation procedure designed to improve computational scalability by separating ROI-level parameter estimation from covariate effect modeling. In the first step, ROI-level hyperparameters are estimated via maximum likelihood. In the second step, we estimate the effects of participant-level covariates on cross-ROI connectivity.
\paragraph{Step 1: Region-by-Region Hyperparameter Estimation}

Ignoring constants, the likelihood for $\mathbf{Z}_{1k}, \cdots, \mathbf{Z}_{Nk}$ in ROI $k$ is
$$
 \mathcal{L}(\btheta_k ; \mathbf{Z}_{1k}, \cdots, \mathbf{Z}_{Nk}) \propto |\bSigma(\btheta_k)|^{-\frac{N}{2}}\exp \Biggl[ - \frac{1}{2}\mbox{tr}\Biggl\{\bSigma(\btheta_k)^{-1}\sum_{i=1}^N\mathbf{Z}_{ik}\mathbf{Z}_{ik}^\top \Biggr\} \Biggr],
$$
% $$
% \sum_{i=1}^N\log \left[(2\pi)^{-n_kT/2}|\bSigma_{{k}}|^{-1/2}\exp\{-\frac{1}{2}\mathbf{Z}_{ik}^\top\bSigma_{z_{ik}}^{-1}\mathbf{Z}_{ik}\}\right],
% $$
where $\mbox{tr}(\cdot)$ is the trace of a matrix and $|\bSigma(\btheta_k)| = \det\{\bSigma(\btheta_k)\}$ denotes the determinant of the covariance matrix. The maximum likelihood estimator (MLE) of $\btheta_k$ based on $(\mathbf{Z}_{ik})_{i=1}^N$ is $\widehat{\btheta}_k= \arg \max_{\btheta_k} \log \mathcal{L}(\btheta_k \mid \mathbf{Z}_{1k}, \cdots, \mathbf{Z}_{Nk})$. If certain parameters (e.g., $\phi$) are assumed common across regions, we form a pooled region-specific estimator $\widehat{\btheta} = (\widehat{\btheta}_1^\text{reg}, \widehat{\btheta}_2^\text{reg}, \widehat{\phi})^\top$, where $\widehat{\phi} = {(\widehat{\phi}_1 + \widehat{\phi}_2)}/{2}$, and $\widehat{\btheta}_k^\text{reg}$ denotes the $\text{ROI}_k$-specific parameters from $\widehat{\btheta}_k$ excluding the shared component $\widehat{\phi}_k$. 

\paragraph{Step 2: Estimation of Covariate Effects on Connectivity}

Using plug-in estimates $\widehat{\lambda}_1, \widehat{\lambda}_2, \bOmega(\widehat{\btheta})$ obtained from Step 1, we estimate the covariate effects $\boldsymbol{\beta}$ by maximizing the log-likelihood under the assumed multivariate normal model due to Theorem \ref{thm:empirical_corr}:
$$
\widehat{\boldsymbol{\beta}} = \arg \max_{\boldsymbol{\beta}} \sum_{i=1}^N \log \mathcal{L}(\mathbf{Y}_i ; \mathbf{X}_i, 
\widehat{\btheta}, \boldsymbol{\beta}),
$$
where $\mathcal{L(\cdot;\cdot)}$ denotes the normal likelihood function under the distribution specified in \eqref{eq:Yi_distribution}.

\subsection{Asymptotic Properties}

Unlike classical geostatistical settings where only a single realization of the process is observed, our framework leverages a large number of independent replicates across participants ($N$). This simplifies estimation of the spatiotemporal parameters as formalized in Proposition \ref{prop:consistency}.

% Proposition \ref{prop:consistency} establishes that $\widehat{\btheta}_k$ is a consistent estimator for $\btheta_{0k}$ when the number of participants approaches infinity, assuming a fixed number of voxels in ROI$_k$ and fixed time points . Intuitively, since $\mathbf{Z}_{ik}$ represents repeated measurements of a spatiotemporal process over a finite support (i.e., a fixed numbers of voxels and time points), consistency of $\widehat{\btheta}_k$ follows from the independency across participants.

\begin{proposition}\label{prop:consistency}
Let $\btheta_{0k}$ denote the true parameters for region $k$. Under Conditions \ref{reg:c1}-\ref{reg:c4}, the MLE $\widehat{\btheta}_k$ defined in Step 1 satisfies $\widehat{\btheta}_k \xrightarrow{p} \btheta_{0k} \quad \text{as } N \to \infty \quad \text{for } k=1,2$.
Let $\widehat{\btheta}$ be the pooled estimator defined in Section \ref{sec:step1}, we have $
\widehat{\btheta} \xrightarrow{p} \btheta_0 \text{ as } N \to \infty$ by the continuous mapping theorem, 
where $\btheta_0$ denotes the true value of the pooled parameter.
\end{proposition}

The second stage of our procedure estimates $\bbeta$ using a plug-in estimator for $\btheta$. Leveraging the consistency of $\widehat{\btheta}$ in Proposition \ref{prop:consistency}, Theorem \ref{thm:asy_beta} establishes the asymptotic properties of the two-stage estimation procedure.

% Theorem \ref{thm:asy_beta} establishes the asymptotic properties of the estimated participant-level covariate effects $\widehat{\bbeta}$ given the consistency of $\widehat{\btheta}$ for plug-in estimation under the M-estimation theory \citep{stefanski2002calculus}. These results justify the use of standard inferential procedures in finite samples, provided the number of participants is sufficiently large.

\begin{theorem}[Asymptotic normality of 
$\widehat{\bbeta}$]\label{thm:asy_beta}
Let $\boldsymbol{\beta}_0$ denote the true regression coefficient vector. Under Conditions \ref{reg:c1}-\ref{reg:c8}, $\widehat{\bbeta}\overset{p}{\rightarrow}\bbeta_0$, i.e., $\widehat{\bbeta}$ is consistent and asymptotically normal as $N\to \infty$ with the limiting distribution $\sqrt{N}(\widehat{\bbeta} - \bbeta_0) \xrightarrow{d} \mathcal{N} \{ \mathbf{0}, \mathbf{I}(\bbeta_0)^{-1} \}$, where $$\mathbf{I}(\bbeta) = \E\left\{
-\frac{\partial^2 \log \mathcal{L}(\mathbf{Y} ; \mathbf{X}, 
{\btheta},{\bbeta})}{\partial \bbeta
\partial \bbeta^\top
}
\right\}.$$
\end{theorem}

Theorem \ref{thm:asy_beta} enables valid large-sample inference and hypothesis testing for $\bbeta$. In finite samples, however, asymptotic variance estimates may be unreliable, and bootstrap methods provide a practical alternative for inference. Accordingly, we estimate the variance of $\widehat{\bbeta}$ using the bootstrap (sampling participants with replacement) in the subsequent simulation and real data analysis.

\section{Simulations}\label{sec:simulation}

We conduct simulation studies to evaluate the performance, comparative advantages, robustness, and practical applicability of the proposed method. The simulations are designed to assess estimation under several settings.

\subsection{Simulation I: Evaluation of Model Performance} \label{sim:baseline}

In the first simulation, we assess the proposed estimation’s performance under scenarios that vary key aspects of the data structure. Specifically, each simulated dataset consists of two brain regions ($k = 1, 2$), with voxel-level signals $Z_{ikt}(\mathbf{s})$ generated from the hierarchical model described in \eqref{eq:model}. 

For each participant $i$, voxel locations are independently sampled from uniform distributions in 3D Euclidean space, with $\boldsymbol{s}_{1j} \sim \text{Unif}([0,1]^3), \boldsymbol{s}_{2j} \sim \text{Unif}([2,3]^3)$ for voxels in ROI$_1$ and ROI$_2$, respectively. Spatial correlation matrices $
r_{s_k}(\mathbf{s},\mathbf{s}')$ are constructed using the exponential kernel $\exp(-\|\mathbf{s} - \mathbf{s}'\|/\psi_k)$, with spatial range parameters $\psi_1 = 1$ and $\psi_2 = 5$. The temporal correlation follows a first-order autoregressive structure $r_\tau(t,t') = \phi^{|t -t'|}$, with $\phi = 0.3$. The spatiotemporal variations $u_{ikt}(\mathbf{s})$ are simulated from a zero-mean Gaussian process with variance components $\lambda_k^2 = 0.4, \sigma_k^2 = 0.3$, and  $\tau_k^2 = 0.3$ for $k = 1,2$. Additionally, each participant is assigned a binary covariate (e.g., autism status), sampled independently from a Bernoulli distribution with equal probability 0.5. The true effects of the intercept and covariate are $\beta_0 = \beta_1 = 0.5$. 

We consider a baseline setting and four variations to test performance with multiple data sizes. The detailed configurations for each scenario are summarized in Table \ref{tab:sim_settings}.
\begin{table}[H]
\centering
\caption{Simulation results across varying scenarios. (a) Simulation settings; (b) Estimation accuracy and coverage; (c) Computation time.}
\label{tab:simulation_combined}

\begin{subtable}[t]{\textwidth}
\centering
\caption{Simulation settings for evaluating model performance.}
\label{tab:sim_settings}
\begin{tabular}{llll}
\toprule
Scenario & Number of Participants & Voxels per Region & Time Points \\
\midrule
Baseline             & 1,000 & 10  & 100 \\
500 Participants   & 500 ($\downarrow$)  & 10  & 100 \\
300 Participants   & 300 ($\downarrow$)  & 10  & 100 \\
Longer Time Series   & 1,000 & 10  & 200 ($\uparrow$)\\
More Voxels          & 1,000 & 30 ($\uparrow$) & 100 \\
\bottomrule
\end{tabular}
\end{subtable}

\vspace{1em}

\begin{subtable}[t]{\textwidth}
\centering
\caption{Step 2 estimation accuracy and uncertainty quantification for $\boldsymbol{\beta} = (\beta_0, \beta_1)$. All but coverage are scaled by \(10^{-3}\).}
\label{tab:beta-estimation}
\begin{tabular}{llrrrrr}
\toprule
Scenario & Parameter & Bias & SE(Bias) & SE & RMSE & Coverage \\
\midrule
\multirow{2}{*}{Baseline}           & \(\beta_0\) & $-0.213$ & 0.718 & 15.9 & 15.9 & 0.946 \\
                                    & \(\beta_1\) &  0.443   & 1.13 & 25.0 & 25.0 & 0.944 \\
\multirow{2}{*}{500 Participants}   & \(\beta_0\) & 0.997    & 1.05 & 22.4 & 22.4 & 0.934 \\
                                    & \(\beta_1\) & 0.737    & 1.70 & 35.4 & 35.4 & 0.932 \\
\multirow{2}{*}{300 Participants}   & \(\beta_0\) & 3.07    & 1.39 & 28.7 & 28.9 & 0.918 \\
                                    & \(\beta_1\) & 3.67    & 2.20 & 45.5 & 45.7 & 0.912 \\
\multirow{2}{*}{Longer Time Series} & \(\beta_0\) & $-0.702$ & 0.532 & 11.1 & 11.2 & 0.935 \\
                                    & \(\beta_1\) & 1.51    & 0.820 & 17.6 & 17.7 & 0.935 \\
\multirow{2}{*}{More Voxels}        & \(\beta_0\) & 0.112    & 0.670 & 15.2 & 15.2 & 0.946 \\
                                    & \(\beta_1\) & 1.48    & 1.17 & 24.1 & 24.1 & 0.928 \\
\bottomrule
\end{tabular}
\end{subtable}

\vspace{1em}

\begin{subtable}[t]{\textwidth}
\centering
\caption{Average computation time (in seconds) for each step across scenarios.}
\label{tab:runtime}
\begin{tabular}{lrrr}
\toprule
Scenario & Step 1 & Step 2 & Total \\
\midrule
Baseline             & 13.9   & 0.197 & 14.2 \\
500 Participants     & 14.5   & 0.124 & 14.6 \\
300 Participants     & 14.7   & 0.100 & 14.8 \\
Longer Time Series   & 59.3   & 0.197 & 59.5 \\
More Voxels          & 161  & 4.41 & 166 \\
\bottomrule
\end{tabular}
\end{subtable}

\end{table}

\noindent Each scenario is independently replicated 500 times. Performance is evaluated using Bias (average bias over 500 replicates), SE(Bias) (standard error of bias), SE (mean of estimated standard errors computed using 100 bootstrap resamples per replicate), root mean squared error (RMSE), and empirical coverage for the 95\% Wald confidence interval using the normal approximation of Theorem \ref{thm:asy_beta} with bootstrapped variance estimate.

A summary of the Step 2 results evaluating across-region covariate effects is presented in Table \ref{tab:beta-estimation}. In the baseline setting, both bias and RMSE are low, and confidence interval coverage is close to the nominal level, suggesting accurate estimation and reliable uncertainty quantification. Performance remains stable under increased temporal or spatial resolution, with longer time series yielding the lowest RMSE. As expected, reducing the number of participants slightly degrades performance, increasing bias and variability, though the model still provides reasonable coverage. Overall, the model demonstrates reliable  performance with different data dimensions.

In addition to statistical accuracy, we report computational efficiency of the proposed method in Table \ref{tab:runtime}. Average runtime is broken down by step: Step 1 (within-region estimation) and Step 2 (across-region covariate effect estimation). As expected, computation time increases with the number of voxels and time points, with voxel-level modeling contributing most significantly to overall cost. Notably, Step 2 is computationally efficient across all scenarios, highlighting the benefits of our model's two-step design.

%\subsection{Simulation II: Comparison with Univariate and Full MLE Methods}

To benchmark the proposed model, we compare its performance against two commonly used alternatives in FC analysis. The first is a univariate ROI-level method that estimates connectivity based on ROI-averaged time series, while the second is a full voxel-level maximum likelihood model that jointly estimates all parameters (full MLE). Details of both competitors are provided in Appendix \ref{appendix:competitors}. As shown in Figure \ref{fig:comparison_plot_baseline}, the proposed method delivers accurate estimation and reliable uncertainty quantification. For both $\beta_0$ and $\beta_1$, it achieves low bias and RMSE comparable to the full MLE method, while maintaining empirical coverage near the nominal 95\%. In contrast, the univariate ROI-averaged approach exhibits inflated bias and RMSE, particularly for $\beta_1$, and tends to underestimate uncertainty. This is partly because it reduces the data to a single average per region, discarding voxel-level information. By using all correlations rather than just the correlation of the average, our method retains more information in the first stage and reduces standard errors. In terms of computational efficiency, the proposed method completes in 14.2 seconds, which is 30 times faster than the full MLE (435.5 seconds) under the same simulation setting. Notably, these benchmarks are based on relatively small datasets; in realistic neuroimaging studies with much larger voxels counts and time series lengths, the full MLE would be prohibitively slow or infeasible. This balance between inferential accuracy and computational scalability highlights the proposed method’s suitability for large-scale neuroimaging studies where voxel-level modeling is desired but full likelihood estimation is computationally infeasible. Additional results under varying sample sizes, time series lengths, and voxel counts (Supplementary Figure \ref{fig:comparison_plot}) further support the efficiency of the proposed method.

\begin{figure}[htbp]
    \centering
    \begin{subfigure}[a]{1\linewidth}
        \centering
        \includegraphics[width=0.7\linewidth]{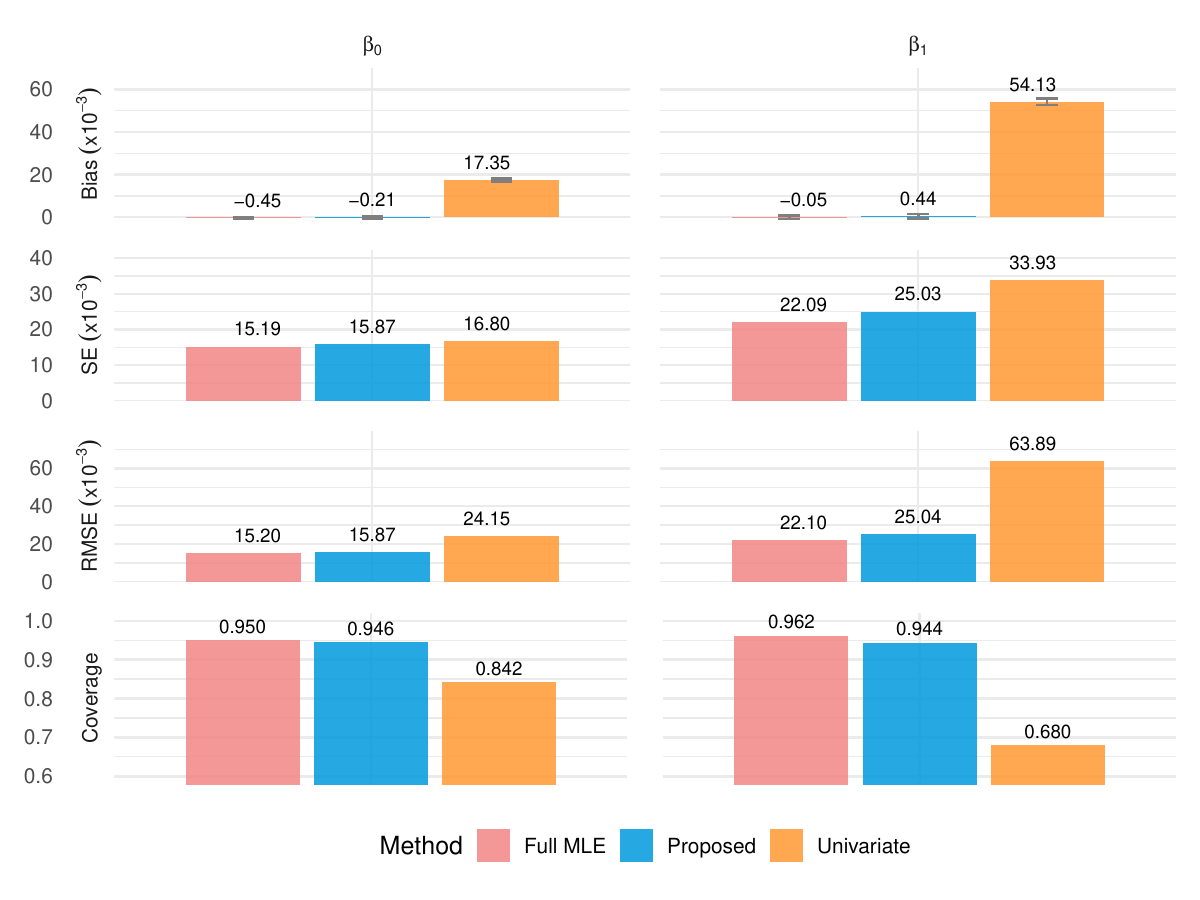}
        \caption{Comparison of estimation accuracy (Bias, SE, RMSE) and coverage probability between the proposed method and two competitors under the baseline simulation scenario. The thin grey vertical segments indicate the associated error bars (±1 SE) for the bias estimates.}
        \label{fig:comparison_plot_baseline}
    \end{subfigure}

    \vspace{1em}

    \begin{subfigure}[b]{0.6\linewidth}
        \centering
        \includegraphics[width=\linewidth]{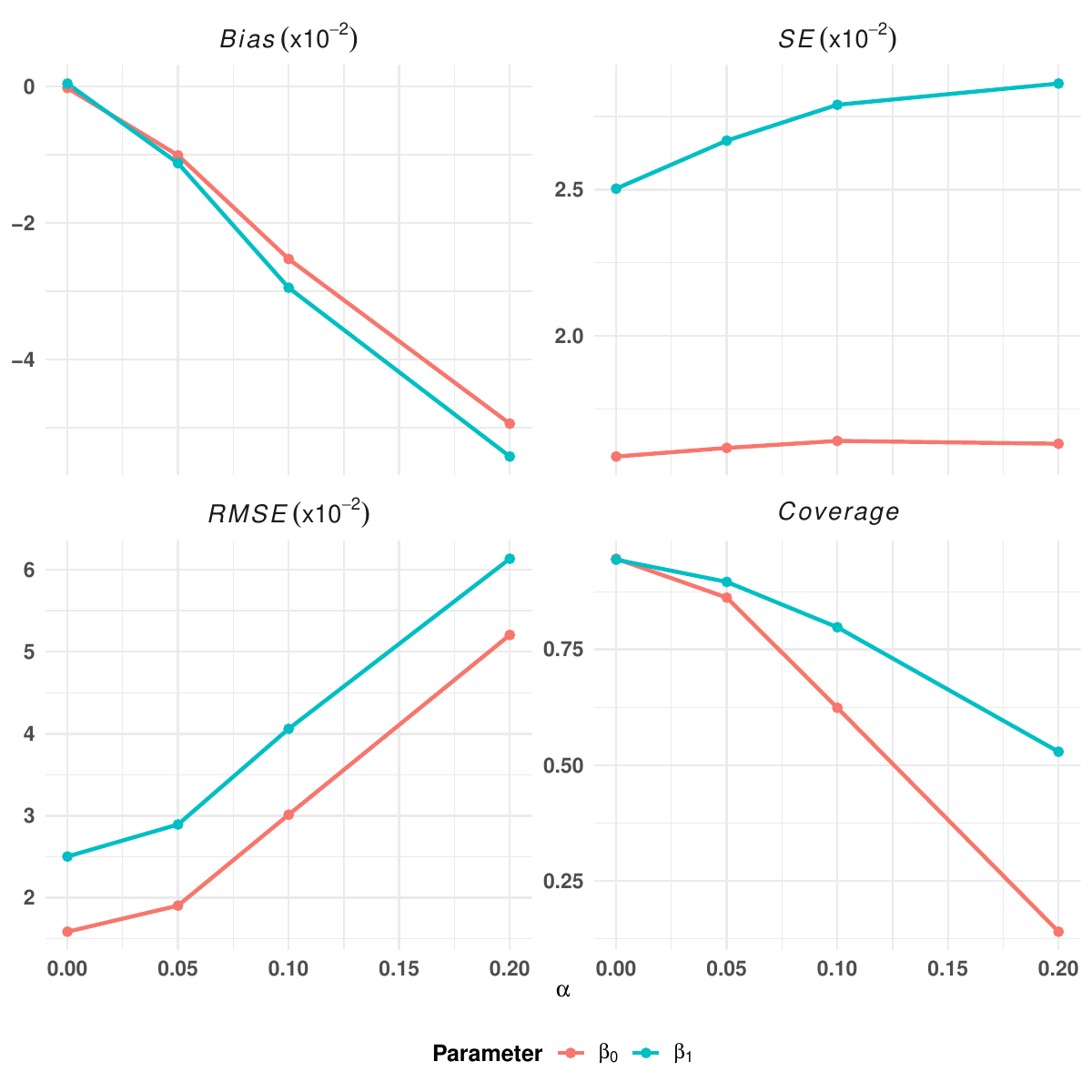}
        \caption{Performance under increasing parameter heterogeneity ($\alpha$), measured by Bias, SE, RMSE (rescaled by $10^{-2}$), and coverage for $\beta_0$ and $\beta_1$, where $\alpha = 0$ represents no voxel-level heterogeneity.}
        \label{fig:robust_plot}
    \end{subfigure}

    \caption{(a) and (b) show estimation and robustness results, respectively.}
    \label{fig:combined_simulation_results}
\end{figure}

\subsection{Simulation II: Robustness to Model Misspecification}\label{sim:robust}

The second simulation investigates the robustness of our model to violations of its underlying assumptions. Specifically, we evaluate the sensitivity of the model's performance when key model parameters, assumed constant across participants in the primary analysis, are instead allowed to vary across participants. Other aspects of the data generation and fitted model are as described in Section \ref{sim:baseline}.

To achieve this, we introduce participant-specific heterogeneity by scaling the baseline scenario parameters such as spatial range parameters $\psi_k$, temporal correlation parameter $\phi$, and variance components $\sigma_k^2$ and $\tau_k^2$ by random factors drawn from a Gamma distribution $\text{Gamma}(1/\alpha,\alpha)$ for each participant. We consider different magnitudes of heterogeneity by varying the scale parameter $\alpha$ across values $\{0.05, 0.1, 0.2\}$, with larger $\alpha$ representing greater parameter variability. Subsequently, we generate voxel-level data using unique spatiotemporal correlation structure and variance patterns for each participant. 

As shown in Figure \ref{fig:robust_plot}, the proposed method performs reliably under mild heterogeneity ($\alpha = 0.05$), with small biases ($< 0.02$) and low RMSEs ($< 0.03$), and empirical coverage above 85\%. As $\alpha$ increases to 0.1 and 0.2, bias and RMSE moderately increase, and coverage declines ---most noticeably for $\beta_0$. However, estimation of $\beta_1$ remains relatively stable, with coverage exceeding 52\% under the most misspecified case. These results indicate that the proposed model is robust to moderate heterogeneity but tends to underestimate uncertainty under more severe heterogeneity. Of note, however, our two-stage method can accommodate subject-specific covariance parameters, as demonstrated in Section \ref{sec:application}, albeit with increased computational cost.

\subsection{Simulation III: Realistic Simulation Based on Real Data}

In the third simulation, we mimic a real-data scenario by generating data that resembles the complexity and structure of the rs-fMRI dataset considered in Section \ref{sec:application}. Specifically, we simulate rs-fMRI outcomes for $n_1=n_2=25$ voxels per region, $T_i=50$ time points, and $N=500$ participants. The variance parameters are identical to those described in  Simulation I. The correlation $\rho_i$ is modeled as a function of an intercept and three covariates: autism status, sampled from a Bernoulli distribution with probability 0.5; standardized age, drawn from a Uniform(0,1) distribution, and gender, sampled from a Bernoulli distribution with probability 0.5. 

As shown in Table \ref{tab:realistic_simulation}, the proposed method accurately recovers the covariate effects across all four regression coefficients. Biases are small, ranging from $0.003$ to $0.006$, RMSEs remain moderate ($0.064$ to $0.106$), and empirical coverage probabilities are close to nominal level, ranging from $0.931$ to $0.944$. The total computation time is 21.8 seconds for Step 1 and 5.4 minutes for Step 2. These results demonstrate that the proposed method generalizes well to larger realistic scenarios, maintaining stable inference in the presence of multiple covariates and a smaller sample size. 

\begin{table}[H]
\centering
\caption{Estimation results for the realistic simulation (Simulation IV). All but coverage are scaled by $10^{-3}$.}
\label{tab:realistic_simulation}
\begin{tabular}{lrrrrr}
\toprule
Parameter & Bias  & SE(Bias) & SE  & RMSE & Coverage  \\
\midrule
$\beta_0$ (Intercept)         & $-1.16$   & 3.09   & 66.9  & 66.9  & 0.940  \\
$\beta_1$ (Autism status)     & $5.94$ & 3.08   & 63.4  & 63.7  & 0.931  \\
$\beta_2$ (Age)               & 0.306   & 5.00   & 106 & 106 & 0.944  \\
$\beta_3$ (Gender)            & $5.35$ & 2.97   & 63.4  & 63.6  & 0.938  \\
\bottomrule
\end{tabular}
\end{table}

\section{Application}\label{sec:application}

\subsection{ABIDE Dataset and Preprocessing}

We use the proposed method to investigate the FC of participants with ASD and neurotypical controls from the Autism Brain Imaging Data Exchange (ABIDE) \citep{di2014autism}, a multi-site study of ASD that aggregated, preprocessed and openly shared rs-fMRI times series data on over $1{,}000$ participants. After excluding participants that did not meet quality control assessments, our analysis includes $238$ individuals with ASD and $300$ neurotypical controls, aged $6$ to $56$ years, who were scanned with a temporal resolution of 2 seconds. 

Preprocessed functional data were obtained from the ABIDE release, processed using the Configurable Pipeline for the Analysis of Connectomes (CPAC) \citep{craddock2013neuro}, which includes motion correction, voxel-wise intensity normalization, and global signal regression. All images were registered to the MNI152 template and resampled to a spatial resolution of $3\times3\times3$ $\text{mm}^3$. To simplify the analysis, for each participant we randomly select $T_i=50$ consecutive time points after discarding the first 10 time points, which are typically excluded to allow for scanner signal stabilization at the beginning of the fMRI acquisition. 

The brain was parcellated using the 200-region Schaefer atlas \citep{schaefer2018local}, which defines ROIs across 17 functional networks. ROI labels were obtained from the corresponding atlas repository \citep{yeoCBIG}. To align the atlas with the fMRI data, the 1 mm Schaefer atlas was downsampled to 3 mm resolution by assigning each $3\times3\times3$ $\text{mm}^3$ voxel the most frequent ROI label within the corresponding high-resolution region. Voxels with non-zero signals across time were retained for analysis, and the signals within each ROI were centered and standardized for each participant.

We focus on connectivity between two ROI pairs selected from three regions across attention-related networks. Details on ROI selection, network labels, and functional interpretations are provided in Section \ref{appendix:real_data_step1}, along with spatiotemporal covariance estimates. 

\subsection{Spatiotemporal Covariance Parameter Estimation}\label{appendix:real_data_step1}

We analyze voxel-wise functional connectivity between two region pairs drawn from three ROIs: the frontal eye fields in the dorsal attention B network, and two distinct subregions within the somatomotor network (SomMotaA$\_11$ and SomMotaA$\_10$), with 28, 32, and 58 voxels respectively. These regions were chosen for their cognitive relevance. The dorsal attention network supports top-down attentional control, while the somatomotor network governs sensorimotor control \citep{corbetta2002control}. Our approach enables quantification of ASD-related alterations in connectivity between attentional and sensorimoter systems while adjusting for other covariates.

\begin{figure}[H]
    \centering
    \includegraphics[width=0.9\linewidth]{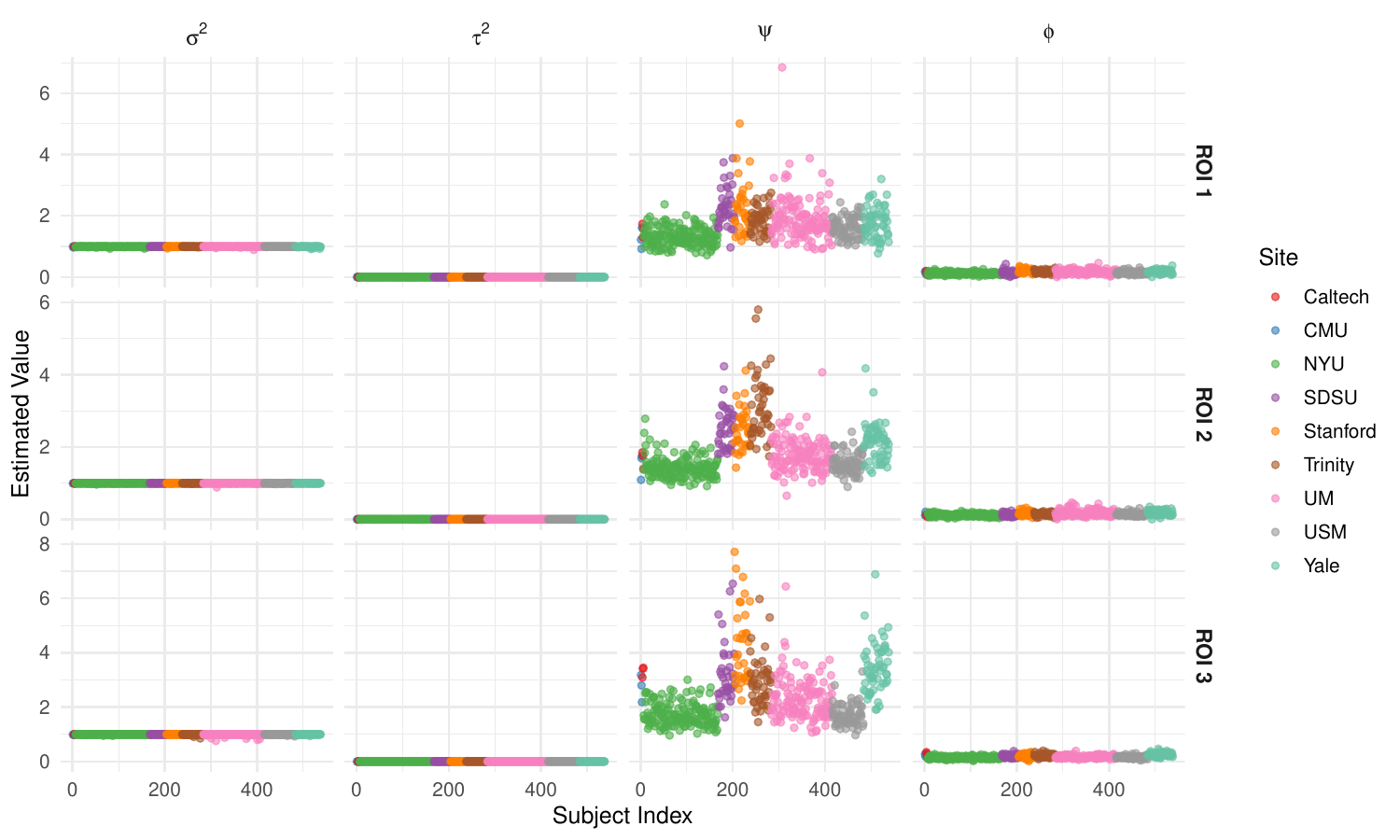}
    \caption{Participant-level estimates of spatiotemporal covariance parameters for each ROI, colored by acquisition site. Each point represents one participant.}
    \label{fig:step1_est_three_ROIs}
\end{figure}

We first estimate voxel-level spatiotemporal covariance parameters --- marginal variance ($\sigma_k^2$), measurement error variance ($\tau_k^2$), spatial range ($\psi_k$), and temporal autocorrelation ($\phi$) --- for each selected ROI. Our simulation study in Section \ref{sim:robust} demonstrates that excessive heterogeneity in these parameters may compromise estimation efficiency and inflate uncertainty. To assess this in our dataset, we first fit Step 1 separately for each participant and visualize the resulting estimates in Figure \ref{fig:step1_est_three_ROIs}. Estimates of $\sigma_k^2, \tau_k^2$ and $\phi$ show minimal variability across individuals, but $\psi_k$ exhibits notable heterogeneity. 
Importantly, we observe that individual estimates of $\psi_k$ tend to cluster by acquisition site, indicating that much of the heterogeneity may stem from site-specific effects. Motivated by this observation, we group participants by site and assume these parameters are homogeneous across participants within each site. This assumption is reasonable given that participants at the same site share scanning protocols, hardware, and preprocessing pipelines. Group-level parameter estimates per site are summarized in Table \ref{tab:covariance_estimates}, and are then used in Step 2 of our estimation procedure. 

\begin{table}[H]
\centering
\caption{Site-specific estimates of spatiotemporal covariance parameters for the three selected ROIs, scaled by $10^{-3}$.}
\label{tab:covariance_estimates}
\begin{tabular}{llrrrr}
\toprule
ROI & Site & $\sigma_k^2$ & $\tau_k^2$ & $\psi_k$ & $\phi$ \\
\midrule
\multirow{9}{*}{Frontal eye fields} 
& Caltech   & 6.90 & 0.992 & 2950 & 158 \\
& CMU       & 7.29 & 0.992 & 2610 & 127 \\
& NYU       & 388 & 0.611 & 1790 & 125 \\
& SDSU      & 172 & 0.827 & 6280 & 150 \\
& Stanford  & 106 & 0.893 & 3820 & 214 \\
& Trinity   & 155 & 0.844 & 3660 & 187 \\
& UM        & 425 & 0.574 & 3340 & 179 \\
& USM       & 196 & 0.803 & 2720 & 150 \\
& Yale      & 151 & 0.848 & 2730 & 193 \\
\midrule
\multirow{9}{*}{Somatomotor A\_11} 
& Caltech   & 969 & 0.0309 & 634000 & 263 \\
& CMU       & 064 & 0.935 & 53600  & 279 \\
& NYU       & 367 & 0.632 & 2160   & 143 \\
& SDSU      & 308 & 0.691 & 15500  & 186 \\
& Stanford  & 570 & 0.430 & 38400  & 204 \\
& Trinity   & 122 & 0.877 & 4230   & 217 \\
& UM        & 289 & 0.711 & 2670   & 169 \\
& USM       & 152 & 0.847 & 2110   & 158 \\
& Yale      & 798 & 0.202 & 27700  & 254 \\
\midrule
\multirow{9}{*}{Somatomotor A\_10} 
& Caltech   & 537 & 2.42 & 14000  & 985 \\
& CMU       & 538 & 2.95 & 12500  & 991 \\
& NYU       & 393 & 0.606 & 1950   & 120 \\
& SDSU      & 569 & 0.431 & 23100  & 131 \\
& Stanford  & 855 & 0.145 & 35800  & 182 \\
& Trinity   & 520 & 0.479 & 19600  & 145 \\
& UM        & 379 & 0.621 & 2810   & 183 \\
& USM       & 159 & 0.840 & 2010   & 145 \\
& Yale      & 219 & 0.780 & 4810   & 188 \\
\bottomrule
\end{tabular}
\end{table}

\subsection{Results}

Using the estimated covariance parameters from Section \ref{appendix:real_data_step1}, we apply the proposed two-step method to estimate associations between voxel-level functional connectivity and covariates consisting of an intercept, autism diagnosis, age, and gender. The analysis is performed separately for the following two ROI pairs: (1) frontal eye fields - somatomotor A$\_11$ and (2) frontal eye fields - somatomotor A$\_10$. For comparison, we also report estimates from the univariate method described in Section \ref{sim:baseline}. 
Table \ref{tab:regression_results} summarizes the estimated regression coefficients and standard errors across both methods. 

For the first pair of ROIs, the proposed method estimates a small negative association between autism and connectivity ($-0.049$, SE $= 0.126$), while the univariate method produces an estimate with a slightly smaller magnitude ($-0.019$, SE $= 0.134$). In contrast, for the second pair of ROIs, the proposed method estimates a larger positive effect of autism on connectivity ($0.127$, SE $= 0.111$), whereas the univariate model reports a smaller but nominally significant effect ($0.040$, SE $= 0.012$). As highlighted in the simulation section, this discrepancy illustrates the impact of a key advantage of the proposed framework: by explicitly modeling voxel-level spatiotemporal covariance, ours is the only approach of the two with calibrated uncertainty quantification. The univariate method, which ignores these dependencies, tends to produce small standard errors and inflates Type-I error rates. These findings are consistent with prior studies studying altered connectivity between the frontoparietal and somatomotor networks in autism, including both decreased connectivity \citep{zhou2024altered} and increased connectivity \citep{rodriguez2025social} respectively.

\begin{table}[H]
\centering
\caption{Estimated regression coefficients and standard errors (in parentheses) for two ROI pairs using the proposed and univariate methods. Asterisks (*) indicate nominal significance at the 0.05 level. Standard errors are computed using 100 nonparametric bootstrap replicates.}
\label{tab:regression_results}
\begin{tabular}{llrr}
\toprule
ROI Pair & Covariate & Proposed & Univariate \\
\midrule
\multirow{4}{*}{Frontal eye fields – Somatomotor A\_{11} } 
& Intercept         & 0.040 (0.089)   & 0.201* (0.104) \\
& Autism            & $-$0.049 (0.126) & $-$0.019 (0.134) \\
& Age               & $-$0.031 (0.143) & 0.050 (0.048) \\
& Gender (female)   & 0.008 (0.121)   & $-$0.236 (0.196) \\
\midrule
\multirow{4}{*}{Frontal eye fields – Somatomotor A\_{10}} 
& Intercept         & 0.136 (0.065)   & 0.102* (0.008) \\
& Autism            & 0.127 (0.111)   & 0.040* (0.012) \\
& Age               & $-$0.022 (0.133) & $-$0.002 (0.006) \\
& Gender (female)   & 0.064 (0.153)   & 0.003 (0.018) \\
\bottomrule
\end{tabular}
\end{table}

\section{Discussion}\label{sec:discussion}

By modeling voxel-wise sample correlations between ROIs and explicitly characterizing their dependence, our approach retains rich voxel-level information that is often lost when reducing the data to a single summary statistic per ROI pair. This enables a more robust assessment of participant-level covariate effects on FC, avoiding underestimated uncertainty and inflated significance that can arise from an oversimplified representation. The proposed two-step estimation procedure further improves computational efficiency by decoupling voxel-level variance modeling from covariate regression, offering a practical alternative to full likelihood methods.

While generalizable, our implementation and focus was restricted to pairwise ROI analysis. We also assumed separability in the spatiotemporal covariance structure, which may be restrictive in some settings. Finally, in finite samples, the plug-in approach in Step 2 may underestimate uncertainty as it does not account for the variability introduced by the Step 1 variance estimation. These limitations suggest opportunities for methodological refinement. Future work includes extending the framework to jointly model connectivity across multiple ROIs or full-brain networks and incorporating joint estimation techniques to better account for first-stage uncertainty. Incorporating longitudinal data or adapting the model to task-based fMRI settings may also broaden its applicability. These extensions will further enhance the framework's utility for individualized and interpretable FC analysis in large neuroimaging studies.

\bibliographystyle{dcu} 
\bibliography{main}

\section*{Supporting Information}

The supplement provides detailed mathematical derivations and additional simulation results that support the main findings of the paper. It includes the proofs of Propositions \ref{prop:pd} and \ref{prop:consistency}, Theorems \ref{thm:empirical_corr} and \ref{thm:asy_beta}, regularity conditions, methodological details for competing methods, and extended simulation performance metrics under varying scenarios. It also reports uncertainty quantification for the univariate method using both analytic and bootstrap-based approaches.

\label{lastpage}

\newpage
\begin{center}
\textbf{Supporting Information for
``Estimating Covariate Effects on Functional Connectivity using Voxel-Level fMRI Data''}
\end{center}
\appendix
\setcounter{table}{0}
\renewcommand{\thetable}{A\arabic{table}}
\setcounter{figure}{0}
\renewcommand{\thefigure}{A\arabic{figure}}

\section{Proof of Theorem \ref{thm:empirical_corr}}\label{appendix:proof_thm1}

We derive the asymptotic distribution of the empirical correlation vector $\mathbf{Y}_i \in \mathbb{R}^{n_1n_2}$, where each element is defined as
$$
Y_i(s_1,s_2) = \frac{1}{T_i}\sum_{t = 1}^{T_i} Z_{i1t}(s_1)Z_{i2t}(s_2), \quad s_1 \in \text{ROI}_1, s_2 \in \text{ROI}_2,
$$
under the spatiotemporal model given in Section \ref{sec:model}. Recall that
$$
Z_{ikt}(s) = a_{ikt} + u_{ikt}(s) + e_{ikt}(s),
$$
where $a_{ikt}$ is the region-level signal shared across voxels in region $k \in \{1,2\}$, $u_{ikt}(s)$ is the voxel-specific spatiotemporal Gaussian processes, and $e_{ikt}(s)$ is the measurement noise. 

We compute the mean and covariance of $Y_i(s_1,s_2)$ using Isserlis' theorem for evaluating expectations of fourth-order moments of Gaussian variables. 

\paragraph{ (A) Mean.}

We first compute the expectation:
$$
\E\{Y_i(s_1,s_2)\} = \E\left\{\frac{1}{T_i}\sum_{t = 1}^{T_i} Z_{i1t}(s_1)Z_{i2t}(s_2)\right\} = \frac{1}{T_i}\sum_{t = 1}^{T_i} \E\{Z_{i1t}(s_1)Z_{i2t}(s_2)\}.
$$
Since $\E\{Z_{ikt}(s)\} = \E(a_{ikt}) + \E\{u_{ikt}(s)\} + \E\{e_{ikt}(s)\} = 0$, $a_{ikt}, u_{ikt}(s)$ and $e_{ikt}(s)$ are mutually independent, and  $u_{ikt}(s), e_{ikt}(s)$ are independent across region, space, and time, the mean of the product $Z_{i1t}(s_1)Z_{i2t}(s_2)$ is determined solely by the covariance between region-level signals:
$$
\E\{Z_{i1t}(s_1)Z_{i2t}(s_2)\} = \Cov(a_{i1t},a_{i2t}) = \rho_i\lambda_1\lambda_2.
$$
Therefore, the expectation of the vector $\mathbf{Y}_i$ is:
$$
\E(\mathbf{Y}_i) = \rho_i\lambda_1\lambda_2\mathbf{1}_{n_1n_2}.
$$

\paragraph{ (B) Covariance.}
The covariance between two connectivity values $Y_i(\mathbf{s}_1,\mathbf{s}_2)$ and $Y_i(\mathbf{s}_3,\mathbf{s}_4)$ is
\begin{align*}
\Cov\left\{ Y_i(\mathbf{s}_1,\mathbf{s}_2), Y_i(\mathbf{s}_3,\mathbf{s}_4) \right\} &= \Cov\left\{\sum_{t_1 = 1}^{T_i} Z_{i1t_1}(\mathbf{s}_1)Z_{i2t_1}(\mathbf{s}_2),\sum_{t_2 = 1}^{T_i} Z_{i1t_2}(\mathbf{s}_3)Z_{i2t_2}(\mathbf{s}_4)\right\}\\
&= \frac{1}{T_i^2} \sum_{t_1,t_2}\Cov\{Z_{i1t_1}(\mathbf{s}_1)Z_{i2t_1}(\mathbf{s}_2),Z_{i1t_2}(\mathbf{s}_3)Z_{i2t_2}(\mathbf{s}_4)\}\\
&= \frac{1}{T_i^2} \sum_{t_1,t_2} [\E\{Z_{i1t_1}(\mathbf{s}_1)Z_{i2t_1}(\mathbf{s}_2)Z_{i1t_2}(\mathbf{s}_3)Z_{i2t_2}(\mathbf{s}_4)\} \\
& \quad - \E\{Z_{i1t_1}(\mathbf{s}_1)Z_{i2t_1}(\mathbf{s}_2)\}\E(\{Z_{i1t_2}(\mathbf{s}_3)Z_{i2t_2}(\mathbf{s}_4)\}].
\end{align*}
Since all components of $Z_{ikt}(\mathbf{s})$ are zero-mean Gaussian and mutually independent across $k,t$, and $\mathbf{s}$, we apply Isserlis's theorem \citep{isserlis1918formula} to expand the 4th-order moment:
\begin{align*}
\E\{Z_{i1t_1}(\mathbf{s}_1)Z_{i2t_1}(\mathbf{s}_2)Z_{i1t_2}(\mathbf{s}_3)Z_{i2t_2}(\mathbf{s}_4)\} &= \E\{Z_{i1t_1}(\mathbf{s}_1)Z_{i2t_1}(\mathbf{s}_2)\}\E\{Z_{i1t_2}(\mathbf{s}_3)Z_{i2t_2}(\mathbf{s}_4)\}\\
&\quad + \E\{Z_{i1t_1}(\mathbf{s}_1)Z_{i1t_2}(\mathbf{s}_3)\}\E\{Z_{i2t_1}(\mathbf{s}_2)Z_{i2t_2}(\mathbf{s}_4)\}\\
&\quad + \E\{Z_{i1t_1}(\mathbf{s}_1)Z_{i2t_2}(\mathbf{s}_4)\}\E\{Z_{i2t_1}(\mathbf{s}_2)Z_{i1t_2}(\mathbf{s}_3)\}.
\end{align*}
Hence, the covariance reduces to 
\begin{align*}
\Cov\left\{ Y_i(\mathbf{s}_1,\mathbf{s}_2), Y_i(\mathbf{s}_3,\mathbf{s}_4) \right\} &= \frac{1}{T_i^2} \sum_{t_1,t_2} [\E\{Z_{i1t_1}(\mathbf{s}_1)Z_{i1t_2}(\mathbf{s}_3)\}\E\{Z_{i2t_1}(\mathbf{s}_2)Z_{i2t_2}(\mathbf{s}_4)\}\\
&\quad + \E\{Z_{i1t_1}(\mathbf{s}_1)Z_{i2t_2}(\mathbf{s}_4)\}\E\{Z_{i2t_1}(\mathbf{s}_2)Z_{i1t_2}(\mathbf{s}_3)\}].
\end{align*}
We compute the second-order moments based on the model 
$$
Z_{ikt}(\mathbf{s}) = a_{ikt} + u_{ikt}(\mathbf{s}) + e_{ikt}(\mathbf{s}),
$$
where $ a_{ikt} \sim \mathcal{N}(0, \lambda_k^2), u_{ikt}(\mathbf{s}) \sim \mathcal{N}(0,\sigma_k^2r_{s_k}(\mathbf{s},\mathbf{s}')r_\tau(t,t'))$, and $e_{ikt}(\mathbf{s}) \sim \mathcal{N}(0,\tau_k^2).$

\emph{Covariance Component Breakdown: }
For $\E\{Z_{i1t_1}(\mathbf{s}_1)Z_{i1t_2}(\mathbf{s}_3)\}$, we get 
\begin{align}\label{eq:mean1}
\E\{Z_{i1t_1}(\mathbf{s}_1)Z_{i1t_2}(\mathbf{s}_3)\} = \lambda_1^2\cdot1(t_1 = t_2) + \sigma_1^2 r_{s_1}(\mathbf{s}_1,\mathbf{s}_3)\phi^{|t_1 - t_2|} + \tau_1^2\cdot1(\mathbf{s}_1 = \mathbf{s}_3, t_1 = t_2).
\end{align}
Similarly for $\E\{Z_{i2t_1}(\mathbf{s}_2)Z_{i2t_2}(\mathbf{s}_4)\}$:
\begin{align}\label{eq:mean2}
\E\{Z_{i2t_1}(\mathbf{s}_2)Z_{i2t_2}(\mathbf{s}_4)\} = \lambda_2^2\cdot1(t_1 = t_2) + \sigma_2^2 r_{s_2}(\mathbf{s}_2,\mathbf{s}_4)\phi^{|t_1 - t_2|} + \tau_2^2\cdot1(\mathbf{s}_2 = \mathbf{s}_4, t_1 = t_2).
\end{align}
The cross-covariance from region-level signals $\E\{Z_{i1t_1}(\mathbf{s}_1)Z_{i2t_2}(\mathbf{s}_4)\}$ and $\E\{Z_{i2t_1}(\mathbf{s}_2)Z_{i1t_2}(\mathbf{s}_3)\}$:
$$
\E\{Z_{i1t_1}(\mathbf{s}_1)Z_{i2t_2}(\mathbf{s}_4)\} = \E\{Z_{i2t_1}(\mathbf{s}_2)Z_{i1t_2}(\mathbf{s}_3)\} = \lambda_1\lambda_2\rho_i\cdot1(t_1 = t_2).
$$

\emph{Covariance matrix form: } To obtain the full covariance matrix of the empirical voxel-wise correlation vector $\mathbf{Y}_i \in \mathbb{R}^{n_1 n_2}$, we aggregate pairwise covariances across all voxel index pairs $(\mathbf{s}_1, \mathbf{s}_2)$, $(\mathbf{s}_3, \mathbf{s}_4)$, where $\mathbf{s}_1, \mathbf{s}_3 \in \mathrm{ROI}_1$ and $\mathbf{s}_2, \mathbf{s}_4 \in \mathrm{ROI}_2$. This results in a covariance matrix of the form:
$$
\bSigma_{\mathbf{Y}_i} = \frac{1}{T_i}\{\lambda_1^2 \lambda_2^2 \rho_i^2 \mathbf{J}_{n_1 n_2} + \bOmega_i(\btheta)\},
$$
The term $\bOmega_i(\btheta)$ contains nine additive components from the product of equation~\eqref{eq:mean1} and \eqref{eq:mean2}:
\begin{align*}
\bOmega_i(\btheta) = &\; 
\lambda_1^2 \lambda_2^2 \mathbf{J}_{n_1 n_2} 
+ \lambda_1^2 \sigma_2^2 (\mathbf{J}_{n_1} \otimes \mathbf{R}_{s_2})
+ \lambda_1^2 \tau_2^2 (\mathbf{J}_{n_1} \otimes \mathbf{I}_{n_2}) \\
&+ \lambda_2^2 \sigma_1^2 (\mathbf{R}_{s_1} \otimes \mathbf{J}_{n_2})
+ \sigma_1^2 \sigma_2^2 m(\phi,T_i) (\mathbf{R}_{s_1} \otimes \mathbf{R}_{s_2})
+ \sigma_1^2 \tau_2^2 (\mathbf{R}_{s_1} \otimes \mathbf{I}_{n_2}) \\
&+ \lambda_2^2 \tau_1^2 (\mathbf{I}_{n_1} \otimes \mathbf{J}_{n_2})
+ \sigma_2^2 \tau_1^2 (\mathbf{I}_{n_1} \otimes \mathbf{R}_{s_2})
+ \tau_1^2 \tau_2^2 (\mathbf{I}_{n_1} \otimes \mathbf{I}_{n_2}),
\end{align*} 
where the temporal autocorrelation is incorporated through a scaling factor
$$m(\phi,T_i) = \frac{1}{T_i}\sum_{t_1=1}^{T_i} \sum_{t_2=1}^{T_i} \phi^{2|t_1 - t_2|} = 1 + \frac{2}{T_i}\sum_{k=1}^{T_i-1}(T_i-k) \phi^{2k} = 1 + 2\sum_{k=1}^{T_i-1}\phi^{2k} - \frac{2}{T_i}\sum_{k=1}^{T_i-1}k\phi^{2k}.$$
Since both finite sums are geometric, we can replace them by their closed-form expressions. For $|\phi| \neq 1$, write $q = \phi^2$:
$$
\sum_{k=1}^{T_i-1}\phi^{2k} = q \frac{1 - q^{T_i-1}}{1-q}, \quad \sum_{k=1}^{T_i-1}k\phi^{2k} = q \frac{1 - T_iq^{T_i-1} + (T_i-1)q^{T_i}}{(1 - q)^2}.
$$
Substituting the closed-from expressions for the geometric sums yields an explicit formula for the scaling factor $m$:
$$
m(\phi,T_i) = 1 + \frac{2\phi^2[1 - \phi^{2(T_i-1)}]}{1 - \phi^2} - \frac{2\phi^2[1 - T_i\phi^{2(T_i-1)} + (T_i-1)\phi^{2T_i}]}{T_i(1-\phi^2)^2},\quad |\phi|<1.
$$
For $|\phi|<1$, as $T_i \rightarrow \infty$, we taking the limit,
$$
\lim_{T_i\to\infty}m(\phi,T_i)
=1+\frac{2\phi^{2}}{1-\phi^{2}} - 0
=\frac{1+\phi^{2}}{1-\phi^{2}}.
$$
In the special case when $|\phi|=1$, the scaling factor reduces to $m(\phi,T_i) = T_i$.

\paragraph{ (C) Proof of Asymptotic Normality.}

    Let $g_{i}(t, \mathbf{s}_1, \mathbf{s}_2) = Z_{i1t}(\mathbf{s}_1) Z_{i2t}(\mathbf{s}_2)$ and define $Y_{i}(\mathbf{s}_1, \mathbf{s}_2) = \sum_{t=1}^{T_i} g_{i}(t, \mathbf{s}_1, \mathbf{s}_2)/T_i$. We show that for any $t_1$ and $\Delta > 0$, the covariance between $g_{i}(t_1, \mathbf{s}_1, \mathbf{s}_2)$ and $g_{i}(t_1+\Delta, \mathbf{s}_1, \mathbf{s}_2)$ is
    \begin{align*}
        \Cov\left\{
        g_{i}(t_1, \mathbf{s}_1, \mathbf{s}_2),
        g_{i}(t_1 + \Delta, \mathbf{s}_1, \mathbf{s}_2)
        \right\} & = 
         \Cov\left\{
        Z_{i1,t_1}(\mathbf{s}_1) Z_{i2, t_1}(\mathbf{s}_2),
        Z_{i1, t_1+\Delta}(\mathbf{s}_1) Z_{i2, t_1+\Delta}(\mathbf{s}_2)
        \right\} \\
        & = \E\left\{
        Z_{i1,t_1}(\mathbf{s}_1) Z_{i2, t_1}(\mathbf{s}_2)Z_{i1, t_1+\Delta}(\mathbf{s}_1) Z_{i2, t_1+\Delta}(\mathbf{s}_2)
        \right\} \\
        &-\E\left\{
        Z_{i1,t_1}(\mathbf{s}_1) Z_{i2, t_1}(\mathbf{s}_2)\right\}
        \E\left\{Z_{i1, t_1+\Delta}(\mathbf{s}_1) Z_{i2, t_1+\Delta}(\mathbf{s}_2)
        \right\} 
    \end{align*}
    Applying Isserlis's theorem \citep{isserlis1918formula} to evaluate the fourth-order moment, the covariance becomes
        \begin{align*}
        \Cov\left\{
        g_{i}(t_1, \mathbf{s}_1, \mathbf{s}_2),
        g_{i}(t_1 + \Delta, \mathbf{s}_1, \mathbf{s}_2)
        \right\} 
        &=\E\left\{
        Z_{i1,t_1}(\mathbf{s}_1) Z_{i1, t_1+\Delta}(\mathbf{s}_1)\right\}
        \E\left\{
        Z_{i2,t_1}(\mathbf{s}_2) Z_{i2, t_1+\Delta}(\mathbf{s}_2)\right\}\\
        &+\E\left\{
        Z_{i1,t_1}(\mathbf{s}_1) Z_{i2, t_1+\Delta}(\mathbf{s}_2)\right\}
        \E\left\{
        Z_{i2,t_1}(\mathbf{s}_2) Z_{i1, t_1+\Delta}(\mathbf{s}_1)\right\}\\
        & = 2\sigma_1^2 \phi^{\Delta},
    \end{align*}
    where $\E\left\{
Z_{i1,t_1}(\mathbf{s}_1) Z_{i2, t_1+\Delta}(\mathbf{s}_2)\right\} = \E\left\{
Z_{i2,t_1}(\mathbf{s}_2) Z_{i1, t_1+\Delta}(\mathbf{s}_1)\right\} = 0$ for $\Delta\neq 0$. Therefore, $g_i(t, \mathbf{s}_1, \mathbf{s}_2)$ follows an AR(1) process for any fixed $\mathbf{s}_1, \mathbf{s}_2$, with $\phi \in (0,1]$, which is strong mixing \citep{bradley2005basic}. By Theorem 27.4, \cite{billingsley2017probability}, the average $Y_{i}(\mathbf{s}_1, \mathbf{s}_2) = \sum_{t=1}^T g_{i}(t, \mathbf{s}_1, \mathbf{s}_2)/T_i$ is asymptotically normal under model \eqref{eq:model}, completing the proof of the asymptotic distribution for $\mathbf{Y}_i$.
\hfill\qedsymbol

\section{Proof of Proposition \ref{prop:pd}}\label{appendix:proof_prop1}

The covariance matrix $\bSigma_{\mathbf{Y}_i} =\{\lambda_1^2 \lambda_2^2 \rho_i^2 \mathbf{J}_{n_1 n_2} + \bOmega_i(\btheta)\}/T_i$ consists of a sum of matrices that are either positive-definite (PD) or positive-semidefinite (PSD). We verify each of the following statements:
\begin{enumerate}[(1)]
    \item The identity matrices $\mathbf{I}_n$ and spatial correlation matrices $\mathbf{R}_{s_1}$ and $\mathbf{R}_{s_2}$ are PD by construction.
    \item The all-ones matrix $\mathbf{J}_n$ is PSD.
    \item Positive scalar multiples of PSD (or PD) matrices are still PSD (or PD).
    \item Kronecker products of PSD (or PD) matrices are still PSD (or PD).
\end{enumerate}
Hence, each term in $\bOmega_i(\btheta)$, as well as the leading term $\lambda_1^2 \lambda_2^2 \rho_i^2 \mathbf{J}_{n_1 n_2}$, is PSD. 

Among the terms in $\bOmega_i(\btheta)$, the component $\sigma_1^2 \sigma_2^2 m(\phi,T_i) (\mathbf{R}_{s_1} \otimes \mathbf{R}_{s_2})$ is PD because both $\sigma_1^2, \sigma_2^2, m(\phi,T_i) > 0$ and $\mathbf{R}_{s_1} \otimes \mathbf{R}_{s_2}$ are PD. Therefore, $\bOmega_i(\btheta)$ contains at least one strictly PD term.

Since the sum of a PD matrix and PSD matrices is PD, $\bSigma_{\mathbf{Y}_i}$ is positive-definite.
\hfill\qedsymbol

\section{Regularity Conditions}

Let $\Theta_k \subset \mathbb{R}^5$ be the compact parametric space of $\btheta_k$. We first list the regularity conditions required to establish large sample properties in this article. 

\begin{enumerate}[C1]
    \item\label{reg:c1}The true parameter $\btheta_{0k}$ lies in the interior of $\Theta_k$.
    \item\label{reg:c2} The participant-specific log-likelihood function $\ell_{ik}(\btheta_k)$ is continuous in $\btheta_k$.
    % \item\label{reg:c3} The expected log-likelihood $\E\{\ell_{ik}(\btheta_k)\}$ is uniquely maximized at $\btheta_{0k}$.
    \item\label{reg:c3} Identifiability: $\btheta_k \neq \widetilde{\btheta}_k$ implies $F(\mathbf{Z}_{ik};\btheta_k) \neq F(\mathbf{Z}_{ik};\widetilde{\btheta}_k)$, where $F(\mathbf{Z}_{ik};\btheta_k)$ denotes the multivariate normal distribution with mean zero and covariance $\bSigma(\btheta_k)$ stated in Section \ref{sec:model}.
    \item\label{reg:c4} The observed log-likelihood satisfies a uniform law of large numbers:
    $$\sup_{\btheta_k \in \Theta_k} \left| \frac{1}{N} \sum_{i=1}^N \ell_{ik}(\btheta_k) - \E\{\ell_{ik}(\btheta_k)\} \right| \xrightarrow{p} 0.
    $$
    \item\label{reg:c5} The estimated covariance matrices $\widehat{\bSigma}_{\mathbf{Y}_i} = \bSigma_{\mathbf{Y}_i}(\widehat{\btheta})$ satisfy $\|\widehat{\bSigma}_{\mathbf{Y}_i} - \bSigma_{\mathbf{Y}_i}\| \xrightarrow{p} 0$ for each $i$.
    \item \label{reg:c6}  For the true regression coefficient vector $\bbeta_0$, the first three partial derivatives of $\log  \mathcal{L}(\mathbf{Y} ; \mathbf{X}, {\btheta}_0, {\bbeta}_0)$ exist in a neighborhood of $\bbeta_0$.
    \item \label{reg:c7} For each $\bbeta$ in a neighborhood of $\bbeta_0$, and for any $i,j,k=1,\cdots,p$, we have
    $$
    \left|\frac{\partial^3 \log  \mathcal{L}(\mathbf{Y} ; \mathbf{X}, 
    {\btheta}_0, {\boldsymbol{\beta}})}{\partial \beta_i \partial \beta_j \partial \beta_k}\right| \leq g(\mathbf{Y}),
    $$
    for all $\mathbf{Y}$ and $\int g(\mathbf{Y})dF(\mathbf{Y})<\infty$.
    \item \label{reg:c8} For each $\bbeta$ in the neighborhood of $\bbeta_0$, we have \begin{align*}
      \mathbf{I}(\bbeta) &= \E\left\{
    -\frac{\partial^2 \log \mathcal{L}(\mathbf{Y} ; \mathbf{X}, 
    {\btheta}_0,{\bbeta}_0)}{\partial \bbeta
    \partial \bbeta^\top
    }
    \right\}\\
    &= 
     \E\left\{
    \frac{\partial \log \mathcal{L}(\mathbf{Y} ; \mathbf{X}, 
    {\btheta}_0,{\bbeta}_0)}{\partial \bbeta
    }
    \frac{\partial \log \mathcal{L}(\mathbf{Y} ; \mathbf{X}, 
    {\btheta}_0,{\bbeta}_0)}{\partial \bbeta^\top
    }
    \right\},
    \end{align*}
    and $\mathbf{I}(\bbeta)$ is non-singular.
\end{enumerate}

All other regularity conditions follow from standard arguments, we therefore now explicitly verify Condition \ref{reg:c3} (Identifiability). For this argument, we require (i) at least two distinct temporal lags (e.g., $T_i \ge 3$); (ii) at least two voxel pairs at distinct distances within ROI$_k$; and (iii) the correlation kernels $r_\tau$ and $r_{s_k}$ are strictly monotone in their parameters. Because the model is Gaussian with zero mean, establishing identifiability reduces to showing that  $\btheta_k \neq \widetilde{\btheta}_k$ implies $\bSigma(\btheta_k) \neq \bSigma(\widetilde{\btheta}_k)$, where $\btheta_k = (\lambda_k^2, \sigma_k^2, \psi_k, \phi, \tau_k^2)^\top$ and $\bSigma(\btheta_k)$ is the covariance matrix of $\bZ_{ik}$. The covariance function is 
$$
\Cov\{Z_{ikt}(\mathbf{s}), Z_{ikt'}(\mathbf{s}')\} = 
\lambda_k^2 \mathbf{1}(t = t') 
+ \sigma_k^2 r_{s_k}(\mathbf{s}, \mathbf{s}') r_\tau(t, t') \, 
+ \tau_k^2 \mathbf{1}(\mathbf{s} = \mathbf{s}', t = t'),
$$
with $r_{s_k}(\mathbf{s}, \mathbf{s}') = \exp(-\|\mathbf{s} - \mathbf{s}'\|/\psi_k)$ and $r_\tau(t, t') = \phi^{|t-t'|}$. 
By scaling, $\lambda_k^2 + \sigma_k^2 + \tau_k^2 = 1$.

Assume two parameter vectors $\btheta_k$ and $\widetilde{\btheta}_k$ generate the same covariance for all $(\mathbf{s},t),(\mathbf{s}',t')$. 
We show that $\btheta_k = \widetilde{\btheta}_k$.

\paragraph{(A) Identify $\phi$ and $\sigma_k^2$ from same-voxel, different-time covariances.}
Fix a voxel $\mathbf{s}$ and consider lags $h = 1,2$ (any two distinct lags suffice). 
For $t \neq t'$,
$$
C^{(\mathrm{sv})}_h := \Cov\{Z_{ikt}(\mathbf{s}), Z_{ik,t+h}(\mathbf{s})\} = \sigma_k^2 \, \phi^{h}.
$$
Equality of covariances under $\btheta_k$ and $\widetilde{\btheta}_k$ yields
$$
\sigma_k^2 \phi = \tilde{\sigma}_k^2 \tilde{\phi}
\quad\text{and}\quad
\sigma_k^2 \phi^{2} = \tilde{\sigma}_k^2 \tilde{\phi}^{2}.
$$
Dividing the second equation by the first gives $\phi = \tilde{\phi}$, and substituting back gives $\sigma_k^2 = \tilde{\sigma}_k^2$.

\paragraph{(B) Identify $\psi_k$ and $\lambda_k^2$ from same-time, different-voxel covariances.}
Fix a time $t$ and choose two voxel pairs with distinct distances $d_1 \neq d_2$. 
For $\mathbf{s} \neq \mathbf{s}'$,
$$
C^{(\mathrm{dv})}(d) := \Cov\{Z_{ikt}(\mathbf{s}), Z_{ikt}(\mathbf{s}')\} 
= \lambda_k^2 + \sigma_k^2 e^{-d/\psi_k}.
$$
From (A), $\sigma_k^2$ is known. Equality of covariances under $\btheta_k$ and $\widetilde{\btheta}_k$ gives
$$
\lambda_k^2 + \sigma_k^2 e^{-d_1/\psi_k} = \tilde{\lambda}_k^2 + \sigma_k^2 e^{-d_1/\tilde{\psi}_k},
$$
$$
\lambda_k^2 + \sigma_k^2 e^{-d_2/\psi_k} = \tilde{\lambda}_k^2 + \sigma_k^2 e^{-d_2/\tilde{\psi}_k}.
$$
Subtracting the second equation from the first yields
$$
\sigma_k^2\left(e^{-d_1/\psi_k} - e^{-d_2/\psi_k}\right)  
= \sigma_k^2\left(e^{-d_1/\tilde{\psi}_k} - e^{-d_2/\tilde{\psi}_k}\right).
$$
Because $d_1 \ne d_2$ and $r_{s_k}(\cdot,\cdot)$ is strictly monotone in $\psi_k$, this implies $\psi_k = \tilde{\psi}_k$.
Substituting back into either equation then gives $\lambda_k^2 = \tilde{\lambda}_k^2$.

\paragraph{(C) Identify $\tau_k^2$ from the variance constraint.}
From $\Var\{Z_{ikt}(\mathbf{s})\} = 1 = \lambda_k^2 + \sigma_k^2 + \tau_k^2$, we obtain $\tau_k^2 = 1 - \lambda_k^2 - \sigma_k^2$, so $\tau_k^2 = \tilde{\tau}_k^2$.

Combining (A)--(C) gives $\btheta_k = \widetilde{\btheta}_k$. Hence, Condition \ref{reg:c3} (identifiability) holds.

\section{Proof of Proposition \ref{prop:consistency}}\label{appendix:proof_consistency}

% Under the assumed model in Section \ref{sec:model}, for each $k \in \{1,2\}$, the vector $\mathbf{Z}_{ik}$ follows a multivariate normal distribution $\mathbf{Z}_{ik} \sim \mathcal{N}(\mathbf{0},\bSigma_{{k}}(\btheta_{0k}))$. The expected log-likelihood is 
% $$
% \E\{l_{ik}(\btheta_k)\} = -\frac{1}{2}\E[n_kT_i\log(2\pi) + \log\{|\bSigma_k(\btheta_k)|\} + \mbox{tr}\{\bSigma_k^{-1}(\btheta_k)\bSigma_k(\btheta_{0k})\}].
% $$
% This expectation is uniquely maximized at the true parameter $\btheta_{0k}$, as it minimizes the Kullback-Leibler divergence between the true distribution $\mathcal{N}(\mathbf{0},\bSigma_{{k}}(\btheta_{0k}))$ and the model $\mathcal{N}(\mathbf{0},\bSigma_{{k}}(\btheta_{k}))$.

Under Conditions \ref{reg:c1}-\ref{reg:c4}, we have 
$$
\widehat{\boldsymbol{\theta}}_k \xrightarrow{p} \boldsymbol{\theta}_{0k}, \quad \text{for } k = 1, 2.
$$
Let \(\widehat{\boldsymbol{\theta}}\) denote the pooled estimator defined in Section \ref{sec:step1}, which is a continuously differentiable function of $\widehat{\boldsymbol{\theta}}_1$ and $\widehat{\boldsymbol{\theta}}_2$. By the continuous mapping theorem, we have 
$$
\widehat{\btheta} \xrightarrow{p} \btheta_0,
$$
where $\btheta_0$ denotes the pooled true parameter. 
\hfill\qedsymbol

\section{Proof of Theorem \ref{thm:asy_beta}}\label{appendix:proof_asy}

The proof is mostly adapted from the proof of Theorem 5.1 in \cite{lehmann2006theory} to our setting. Denote $\ell(\widehat{\btheta},\widehat{\boldsymbol{\beta}})=\sum_{i=1}^N \log  \mathcal{L}(\mathbf{Y}_i ; \mathbf{X}_i, 
\widehat{\btheta}, \widehat{\boldsymbol{\beta}})$. Under Conditions \ref{reg:c1}-\ref{reg:c4}  and continuous mapping theorem, we have $\widehat{\bbeta}\overset{p}{\rightarrow}\bbeta_0$, i.e., $\widehat{\bbeta}$ is consistent. 

By the Taylor series expansion of $\partial \sum_{i=1}^N \log  \mathcal{L}(\mathbf{Y}_i ; \mathbf{X}_i, 
\widehat{\btheta}, \widehat{\boldsymbol{\beta}})/\partial \boldsymbol{\beta}$ at the true regression coefficient vector $\boldsymbol{\beta}_0$, we have
\begin{equation*}
    \boldsymbol{0}=\frac{\partial \ell(\widehat{\btheta},\widehat{\bbeta})}{\partial \bbeta} = 
\frac{\partial \ell(\widehat{\btheta},{\bbeta_0})}{\partial \bbeta} + 
\frac{\partial^2 \ell(\widehat{\btheta},{\bbeta_0})}{\partial \bbeta\partial \bbeta^\top} (\widehat{\bbeta}- \bbeta_0) + \frac{1}{2}
\left(\sum_{j,k}\frac{\partial^3 \ell(\widehat{\btheta},\bbeta^*)}{\partial \beta_i \partial \beta_j \partial \beta_k} (\widehat{\beta}_j - \beta_{j0})(\widehat{\beta}_{k} - \beta_{k0})\right)^{p}_{i=1},
\end{equation*}
where $\bbeta^*$ is a point on the line segment connecting $\widehat{\bbeta}$ and $\bbeta$, and the left side is zero by the definition of maximizer. Therefore, the $i$-th entry of the equation can be written as
$$
\sum_{j=1}^p\sqrt{N}
(\widehat{\beta}_j - \beta_{j0})
\left\{
\frac{1}{N}
\frac{\partial^2l(\widehat{\btheta},{\bbeta}_0)}{\partial \beta_i\partial \beta_j} +
\frac{1}{2N}
\sum_{k}\frac{\partial^3 \ell(\widehat{\btheta},\bbeta^*)}{\partial \beta_i \partial \beta_j \partial \beta_k} (\widehat{\beta}_{k} - \beta_{k0})
\right\} = -\frac{1}{\sqrt{N}}\frac{\partial \ell(\widehat{\btheta},{\bbeta_0})}{\partial \beta_i}
$$
for $i=1,\cdots,p$. 
Under Conditions \ref{reg:c5}-\ref{reg:c7}, we have 
$$
\frac{1}{N}\frac{\partial^2l(\widehat{\btheta},{\bbeta}_0)}{\partial \beta_i\partial \beta_j} +
\frac{1}{2N}
\sum_{j,k}\frac{\partial^3 \ell(\widehat{\btheta},\boldsymbol{\beta}^*)}{\partial \beta_i \partial \beta_j \partial \beta_k} (\widehat{\beta}_{k} - \beta_{k0})
=
\left[\E\left\{
\frac{\partial^2 \log \mathcal{L}(\mathbf{Y} ; \mathbf{X}, 
{\btheta}_0,{\boldsymbol{\beta}}_0)}{\partial \boldsymbol{\beta} \partial \boldsymbol{\beta}^\top}
\right\}\right]_{i,j} + O_p(N^{-1}),
$$
as $\widehat{\boldsymbol{\theta}} \xrightarrow{p} \boldsymbol{\theta}_0$ by Proposition \ref{prop:consistency} and continuous mapping theorem. Under Condition \ref{reg:c8}, the limiting distribution of $\sqrt{N}(\widehat{\beta}_j - \beta_{j0})$ is 
\begin{align*}
    \sum_{j=1}^p\sqrt{N}(\widehat{\beta}_j - \beta_{j0})\left[
\E\left\{
\frac{\partial^2 \log \mathcal{L}(\mathbf{Y} ; \mathbf{X}, 
{\btheta}_0,{\bbeta}_0)}{\partial \bbeta \partial \bbeta^\top}
\right\}
\right]_{i,j}&=-\frac{1}{\sqrt{N}} \frac{\partial \ell({\btheta}_0,{\bbeta_0})}{\partial {\beta}_i} + o_p(1),
\end{align*}
where ${N}^{-1/2}\{{\partial \ell({\btheta}_0,{\bbeta_0})}/{\partial {\beta}_i}\}_{i=1}^p$ converges in distribution to a multivariate normal random variable with mean zero and covariance matrix $\mathbf{I}(\bbeta_0)$ under Conditions \ref{reg:c5}-\ref{reg:c7}. Hence, we have
\begin{align*}
    \sqrt{N}(\widehat{\bbeta} - \bbeta_0)&=-\frac{1}{\sqrt{N}} \left[
\E\left\{
\frac{\partial^2 \log \mathcal{L}(\mathbf{Y} ; \mathbf{X}, 
{\btheta}_0,{\bbeta}_0)}{\partial \bbeta \partial \bbeta^\top}
\right\}
\right]^{-1}\frac{\partial \ell({\btheta}_0,{\bbeta_0})}{\partial \bbeta} + o_p(1)\\
&\overset{d}{\rightarrow} \mathcal{N}\{\mathbf{0}, \mathbf{I}(\bbeta_0)^{-1}\},
\end{align*}
which completes the proof.
\hfill\qedsymbol

\section{Competing Methods}

\subsection{Methodological Details}\label{appendix:competitors}

This section provides details for the two competing methods used in the simulation comparison. 

Univariate method is a region-level analysis that averages voxel-wise signals within each ROI. For participant $i$ and region $k$, the averaged signal is computed as 
$$
Z_{ikt}^{(\text{avg})} = \frac{1}{n_k}\sum_{s\in \text{ROI}_k} Z_{ikt}(\mathbf{s}).
$$
Participant-specific connectivity $\rho_i$ is then estimated as the Pearson correlation between the two regional averages,  $$
\rho_i = \frac{\Cov(Z_{i1t}^{(\text{avg})}, Z_{i2t}^{(\text{avg})})}{\lambda_1\lambda_2}.
$$ 
Let $\mathbf{X} \in \mathbb{R}^{N\times p}$ denote the design matrix. After applying the logit transformation to the correlations, we estimate the regression coefficients using ordinary least squares:
$$
\widehat{\bbeta} = (\mathbf{X}^\top \mathbf{X})^{-1}\mathbf{X}^\top \text{logit}\left(\frac{\rho_i+1}{2}\right)
$$
We quantify uncertainty in the estimated covariate effects using a nonparametric bootstrap approach, resampling participants and recomputing estimates over 100 datasets. We also use the closed-from variance estimate from the linear model:
$$
\widehat{\Var}(\widehat{\bbeta}) = \widehat{\sigma}^2(\mathbf{X}^\top \mathbf{X})^{-1},
$$
where $\widehat{\sigma}^2 = \|\text{logit}\{(\rho_i + 1)/2\} - \mathbf{X}\widehat{\bbeta}\|/(N-p)$ is the residual variance estimate. The closed-form results are provided in Table \ref{tab:comp1_results}. By averaging voxel-level signals within each ROI, the univariate approach effectively treats each region as homogeneous, ignoring voxel-level spatial variability. While this aggregation simplifies estimation and improves computational efficiency, it disgards within-region heterogeneity, which may lead to underestimation of uncertainty and biased inference in finite samples. 

Another method is a full voxel-level maximum likelihood method (Full MLE) that jointly models the high-dimensional spatiotemporal signals across all voxels and regions. Its likelihood function is expressed as 
$$
\mathcal{L}(\bbeta,\btheta) = \prod_{i=1}^N\mathcal{N}(\mathbf{Z}_i; \mathbf{X}_i \bbeta, \bSigma_i(\btheta)),
$$
where $\bSigma_i(\btheta)$ is a covariance matrix of dimension $(n_1 + n_2) T \times (n_1 + n_2) T$. The model parameters including regression coefficients, spatiotemporal decay parameters, and variance components are estimated simultaneously by maximizing the joint log-likelihood over all participants. While accounting for voxel-level structure and uncertainty, this method increases computational cost substantially due to repeated inversion of large covariance matrices. Due to the substantial computational burden of Full MLE, we only implement this method under the baseline simulation scenario. 

\subsection{Simulation Performance under Varying Scenarios}\label{appendix:additional_simulation}

To evaluate the scalability of the proposed method, we conduct additional simulation studies under alternative scenarios. The baseline scenario is included for reference. 

\begin{figure}[H]
    \centering
    \includegraphics[width=1\linewidth]{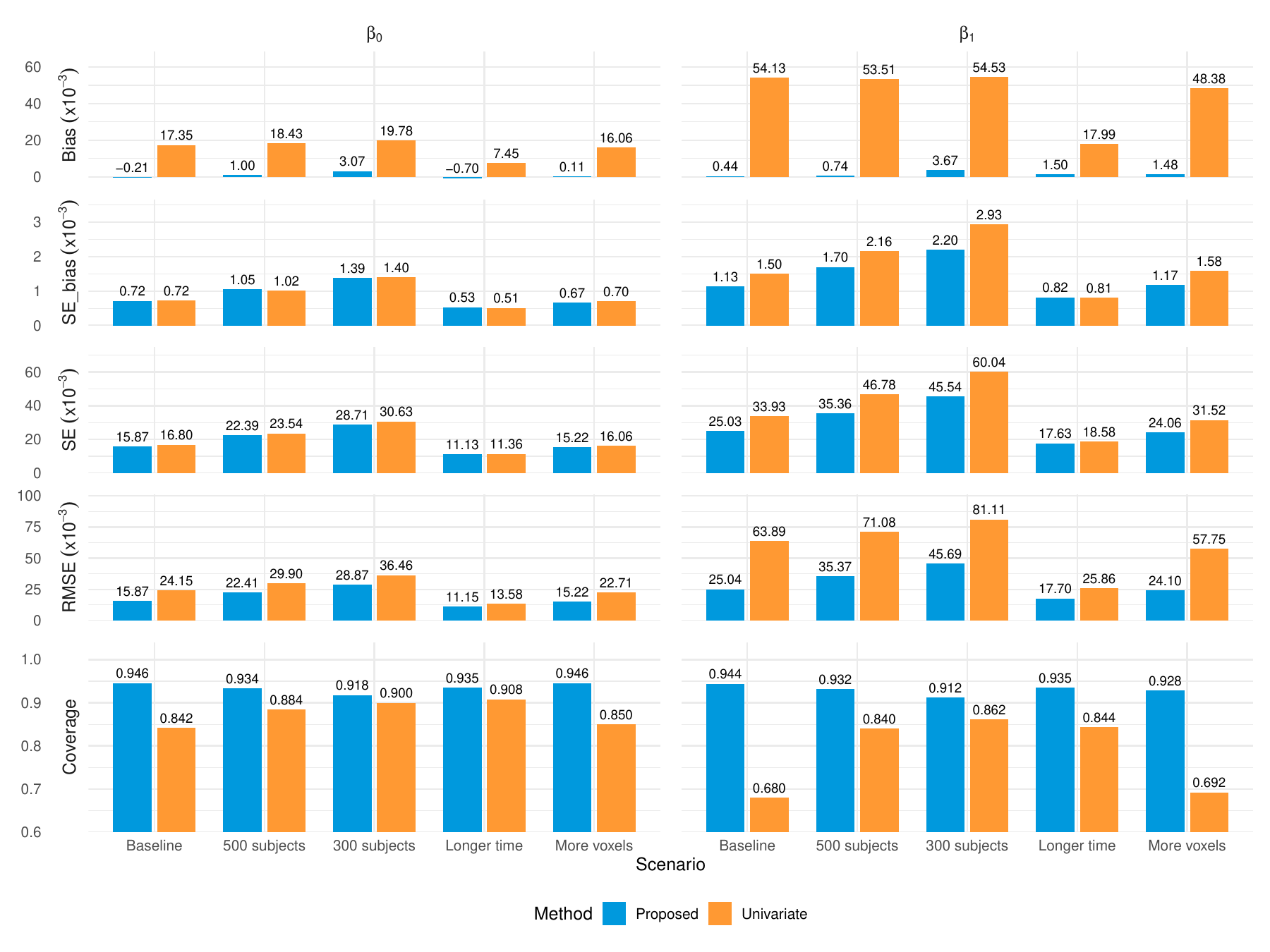}
    \caption{Comparison of estimation accuracy (Bias, SE, RMSE) and coverage probability between the proposed method and Univariate method across simulation scenarios. The baseline scenario is repeated from the main text for comparison. The thin grey vertical segments indicate the associated error bars (±1 SE) for the bias estimates.}
    \label{fig:comparison_plot}
\end{figure}

Figure \ref{fig:comparison_plot} presents the estimation performance of the proposed method and Univariate method across all simulation scenarios. The proposed method maintains low bias, small SE and RMSE, and near-nominal coverage across varying data configurations. Notably, performance remains stable even when the number of voxels increases. In contrast, Univariate method consistently shows larger bias and under-coverage, particularly for $\beta_2$, with its performance further deteriorating in the fewer participants and higher dimensional scenarios. 

\subsection{Uncertainty Quantification for Univariate Method}\label{appendix:competitor1}

In this section, we report the analytic variance estimates for Univariate method and comparison to bootstrap based uncertainty quantification. 

\begin{table}[H]
\centering
\caption{Estimation accuracy and uncertainty quantification for the univariate method under various simulation scenarios. Bias, SE, and RMSE are in $10^{-3}$. Results are shown for both analytic estimates and bootstrap-based estimates (SE$_{\text{boot}}$, RMSE$_{\text{boot}}$, Coverage$_{\text{boot}}$).}
\label{tab:comp1_results}
\setlength{\tabcolsep}{4pt}
\begin{tabular}{llrrrrrrr}
\toprule
Scenario & Parameter & Bias & SE & SE$_{\text{boot}}$ & RMSE  & RMSE$_{\text{boot}}$ & Coverage & Coverage$_{\text{boot}}$ \\
\midrule
Baseline             & $\beta_0$ & 17.3 & 24.1 & 16.8 & 29.7 & 24.2 & 0.962    & 0.842 \\
                     & $\beta_1$ & 54.1 & 34.0 & 33.9 & 63.9 & 63.9 & 0.690   & 0.680 \\\\
500 Participants     & $\beta_0$ & 18.4 & 33.2 & 23.5 & 38.0 & 29.9 & 0.948    & 0.884 \\
                     & $\beta_1$ & 53.5 & 47.0 & 46.8 & 71.2 & 71.1 & 0.846    & 0.840 \\\\
300 Participants     & $\beta_0$ & 19.8 & 42.7 & 30.6 & 47.0 & 36.5 & 0.962    & 0.900 \\
                     & $\beta_1$ & 54.5 & 60.3 & 60.0 & 81.3 & 81.1 & 0.860    & 0.862 \\\\
Longer time series   & $\beta_0$ & 7.40 & 13.2 & 11.4 & 15.2 & 13.6 & 0.946    & 0.908 \\
                     & $\beta_1$ & 18.0 & 18.7 & 18.6 & 25.9 & 25.9 & 0.842    & 0.844 \\\\
More voxels          & $\beta_0$ & 16.1 & 22.4 & 16.1 & 27.5 & 22.7 & 0.942    & 0.850 \\
                     & $\beta_1$ & 48.4 & 31.6 & 31.5 & 57.8 & 57.7 & 0.692   & 0.692 \\
\bottomrule
\end{tabular}
\end{table}
\end{document}